\documentclass[acmsmall,screen,authorversion,nonacm]{acmart}

\usepackage{fancyhdr}
\AtBeginDocument{%
    \addtolength{\footskip}{2.0\baselineskip}%
    \fancyfoot[L]{\textit{\textbf{To appear in SC '21}}}%
}

\usepackage{balance}
\usepackage{mdwmath}
\usepackage{mdwtab}
\usepackage{graphicx}
\usepackage[tight,footnotesize]{subfigure}
\usepackage{xspace}
\usepackage{listings}
\usepackage{stfloats}
\usepackage{float,rotating}
\usepackage{flushend}
\usepackage{url}
\usepackage{xcolor}

\usepackage{enumitem}
\usepackage{amsmath}
\usepackage{adjustbox}
\usepackage[linesnumbered,ruled, lined]{algorithm2e}
\usepackage{multirow}
\usepackage[noend]{algpseudocode}
\usepackage{xcolor}
\usepackage{xhfill}
\begin{document}


\newcommand{\ditto}[1][.6pt]{\xrfill[.5ex]{#1}~\textquotedbl~\xrfill[.5ex]{#1}}
\newcommand{\fix}[1]{\textcolor{red}{#1}}
\newcommand{\rebuttal}[1]{{#1}}
\newcommand{\reply}[1]{\textcolor{blue}{#1}}
\newcommand{\doubt}[1]{\textcolor{yellow}{\textbf{#1}}}
\newcommand{\sridutt}[1]{\textcolor{violet}{#1}}

\newenvironment{algocolor}{%
   \setlength{\parindent}{0pt}
}{}
\newcommand{\cfl}[1]{\textjava{CF}{\textsubscript{\textjava{#1}}}\xspace}
\newcommand{\ufl}[1]{\textjava{UF}{\textsubscript{\textjava{#1}}}\xspace}

\newcommand{\jpi}[1]{JPI{\textsubscript{#1}}}
\newcommand{\fl}[1]{FQ{\textsubscript{#1}}}
\newcommand{\nprev}{N{\textsubscript{prev}}}
\newcommand{\ncurr}{N{\textsubscript{curr}}}
\newcommand{\cfa}[1]{CF{\textsubscript{#1}}}
\newcommand{\ufa}[1]{UF{\textsubscript{#1}}}
\newcommand{\tuf}{{NUM\textsubscript{UF}}\xspace}
\newcommand{\tcf}{{NUM\textsubscript{CF}}\xspace}
\newcommand{\range}{Range\xspace}
\newcommand{\ratio}{\alpha\xspace}

\newcommand{\clb}{{CF\textsubscript{LB}}\xspace}
\newcommand{\ulb}{{UF\textsubscript{LB}}\xspace}
\newcommand{\crb}{{CF\textsubscript{RB}}\xspace}
\newcommand{\urb}{{UF\textsubscript{RB}}\xspace}
\newcommand{\oufest}{{UF\textsubscript{opt\_est}}\xspace}
\newcommand{\ocf}{{CF\textsubscript{opt}}\xspace}
\newcommand{\ouf}{{UF\textsubscript{opt}}\xspace}
\newcommand{\cfstart}{\textjava{cuttlefish::start()}\xspace}
\newcommand{\cfstop}{\textjava{cuttlefish::stop()}\xspace}
\newcommand{\omp}{{OpenMP}\xspace}
\newcommand{\fmaxmax}{{CF\textsubscript{max}\_UF\textsubscript{max}}\xspace}
\newcommand{\fmidmax}{{CF\textsubscript{mid}\_UF\textsubscript{max}}\xspace}
\newcommand{\fmaxmid}{{CF\textsubscript{max}\_UF\textsubscript{mid}}\xspace}
\newcommand{\fminmax}{{CF\textsubscript{min}\_UF\textsubscript{max}}\xspace}
\newcommand{\fmaxmin}{{CF\textsubscript{max}\_UF\textsubscript{min}}\xspace}
\newcommand{\interval}{{T\textsubscript{inv}}\xspace}
\newcommand{\torins}{\textjava{TOR_INSERT}\xspace}
\newcommand{\hc}{{HClib}\xspace}
\newcommand{\uts}{\textjava{UTS}\xspace}
\newcommand{\sorws}{\textjava{SOR-ws}\xspace}
\newcommand{\heatws}{\textjava{Heat-ws}\xspace}
\newcommand{\sorrr}{\textjava{SOR-rt}\xspace}
\newcommand{\heatrr}{\textjava{Heat-rt}\xspace}
\newcommand{\sorir}{\textjava{SOR-irt}\xspace}
\newcommand{\heatir}{\textjava{Heat-irt}\xspace}
\newcommand{\minife}{\textjava{MiniFE}\xspace}
\newcommand{\hpccg}{\textjava{HPCCG}\xspace}
\newcommand{\amg}{\textjava{AMG}\xspace}

\newcommand{\cf}{{Cuttlefish}\xspace}
\newcommand{\cfc}{{Cuttlefish-Core}\xspace}
\newcommand{\cfu}{{Cuttlefish-Uncore}\xspace}
\newcommand{\dtp}{{DTP}\xspace}
\newcommand{\ab}{\textjava{Amoeba}\xspace}
\newcommand{\abs}{\textjava{AmoebaSequential}\xspace}
\newcommand{\abst}{\textjava{AmoebaStatic}\xspace}
\newcommand{\orig}{\textjava{Default}\xspace}
\newcommand{\origseq}{\textjava{DefaultSequential}\xspace}
\newcommand{\seq}{\textjava{Sequential}\xspace}
\newcommand{\cilk}{\textjava{CilkPlus}\xspace}
\newcommand{\finish}{\textjava{finish}\xspace}
\newcommand{\async}{\textjava{async}\xspace}
\newcommand{\spawn}{\textjava{spawn}\xspace}
\newcommand{\sync}{\textjava{sync}\xspace}
\newcommand{\fasync}{\textjava{forasync1D}\xspace}
\newcommand{\afp}{\async--\finish\xspace}
\newcommand{\sps}{\spawn--\sync\xspace}
\newcommand{\pluseq}{\mathrel{+}=}
\newcommand{\mineq}{\mathrel{-}=}
\newcommand{\pinterval}{\textjava{P_INTERVAL}\xspace}


\newcommand*{\textbmx}[1]{#1}  
\newcommand*{\textbm}[1]{\textbmx{#1}\xspace}

\newcommand{\RemoveSpaces}[1]{%
  \begingroup
  \spaceskip=1sp
  \xspaceskip=1sp
  #1%
  \endgroup}



\newcommand{\doi}[1]{doi:~\href{http://dx.doi.org/#1}{\Hurl{#1}}}

%
%

\newcommand{\eg}{e.g., }
\newcommand{\ie}{i.e., }
\newcommand{\etal}{et al. }

%

\lstloadlanguages{Java}
\lstset{
  numbers=left,
  numberstyle=\tiny\sffamily,
  stepnumber=1,
  numbersep=1em,
  language=java,                         
  basicstyle=\scriptsize\ttfamily,     
  commentstyle=\scriptsize\it\ttfamily,
  stringstyle=\ttfamily,
  escapechar={\$},
  morekeywords={async, launch, finish, forasync, spawn, sync, loop\_domain\_t, forasync1D}
}

\DeclareRobustCommand{\textjava}[1]{{\lstset{basicstyle=\ttfamily}\lstinline@#1@}}
\newcommand{\cnull}{\textjava{NULL}\xspace}



\title{Cuttlefish: Library for Achieving Energy Efficiency in Multicore Parallel Programs}

\author{Sunil Kumar}
\affiliation{
  \institution{IIIT-Delhi}
  \country{India}
}
\author{Akshat Gupta}
\affiliation{
  \institution{IIIT-Delhi}
  \country{India}
}
\author{Vivek Kumar}
\affiliation{
  \institution{IIIT-Delhi}
  \country{India}
}
\author{Sridutt Bhalachandra}
\affiliation{
  \institution{Lawrence Berkeley National Laboratory}
  \country{USA}
}

\renewcommand{\shortauthors}{S. Kumar et al.}


\begin{abstract}

A low-cap power budget is challenging for exascale computing. Dynamic Voltage and Frequency Scaling (DVFS) and Uncore Frequency Scaling (UFS) are the two widely used techniques for limiting the HPC application's energy footprint. However, existing approaches fail to provide a unified solution that can work with different types of parallel programming models and applications. 

This paper proposes \emph{\cf}, a programming model oblivious C/C++ library for achieving energy efficiency in multicore parallel programs running over Intel processors. An online profiler periodically profiles model-specific registers to discover a running application's memory access pattern. Using a combination of DVFS and UFS, \cf then dynamically \emph{adapts} the processor's core and uncore frequencies, thereby improving its energy efficiency. The evaluation on a 20-core Intel Xeon processor using a set of widely used OpenMP benchmarks, consisting of several irregular-tasking and work-sharing pragmas, achieves geometric mean energy savings of 19.4\% with a 3.6\% slowdown.

\end{abstract}



\maketitle

\section{Introduction}
\label{sec:intro}

The current generation of supercomputers is composed of highly multicore processors.
The ever-increasing core counts are likely to continue in the present and post-Exascale ($10^{18}$ flops/sec) era.
The Top500 list for November 2020 shows 50\% of supercomputers have 20-24 cores per socket, and the fastest supercomputer, Fugaku, has 48 cores~\cite{top500}.
Limiting power consumption is one of the significant challenges today.
For an exascale system to be delivered within a power budget of 20–30 MW~\cite{exascale}, the hardware and software components need to be highly energy-efficient.

Cache-coherent shared memory processors that dominate modern high-performance computing (HPC) systems contain multiple levels of caches in each socket.
Most processor architectures are a combination of \emph{core} and \emph{uncore} elements.
The uncore includes all chip components outside the CPU core~\cite{ufs}, such as shared caches, memory controllers, and interconnects (QPI on Intel platforms).
The processor power consumption can be regulated using
 multiple knobs for such as Dynamic Voltage and Frequency Scaling (DVFS)~\cite{kim2008system} and Dynamic Duty-Cycle Modulation (DDCM)~\cite{ddcm} for CPU cores, and Uncore Frequency Scaling (UFS)~\cite{ufs} for the uncore.
DVFS allows scaling down the core's voltage and frequency, thereby reducing
power consumption as lowering the
voltage has a squared effect on active power consumption.
 The static losses have become significant with transistor shrinkage, undermining the 
previously achievable power savings with DVFS~\cite{esmaeilzadeh2011dark}.
 Nevertheless, DVFS is still the most effective and widely used power control.

Most energy-efficient HPC research has revolved around reducing the processor/core frequency by using DVFS or DDCM  with minimum performance impact (see Section~\ref{sec:related}).
These implementations adapt the core frequencies as follows: 
 a) based on trace data collected from the application's offline profiling,
 b) whenever parallel programs encounter slack time due to unbalanced computation or inter-node communication delays,
 c) gathering workload characteristics by profiling iterative and loop-based applications, and
 d) programming model and runtime-dependent application profiling.
 A standard limitation of these studies is that they are specific to a particular application or programming model.
 Recently, UFS has been explored for achieving energy efficiency based on machine learning models on offline generated application data~\cite{BSAB2019} and by dynamically profiling the DRAM power consumption~\cite{GNMF2019}.
 However, these studies lack an integrated approach for core and uncore frequency scaling.

This paper explores a one-stop solution
for achieving energy efficiency on multicore processors
called \cf that dynamically adapts both the core and uncore
frequencies on Intel processors without requiring training runs. \cf is oblivious
to the parallel programming model and the concurrency decomposition
techniques used in an application. 
In their application, programmers only need
to use two API \rebuttal{functions}, \cfstart and \cfstop, to
define the scope requiring an energy-efficient execution.
At runtime, \cf then creates a daemon thread that periodically
profiles the application's Memory Access Pattern (MAP)
by reading the Model-Specific Registers (MSR) available
on all Intel platforms. Each MAP is uniquely
identified as the ratio of uncore event \torins~\cite{torinsert}
and instructions retired.
\cf daemon then uses DVFS to 
determine the core frequency that would provide
maximum energy savings with minimal impact on execution time.
After finding and setting this optimal core frequency, \cf daemon then uses the
same exploration-based technique to determine the optimal uncore frequency.
To reduce the loss in performance, it uses several runtime optimizations to complete 
the frequency explorations quickly.
Once both optimal core and uncore frequencies are found for 
a given MAP, the rest of the program would
execute at these frequencies. \cf would repeat
this frequency exploration for core and uncore
every time a new MAP is discovered. We chose a set of
ten widely-used \omp benchmarks, consisting of several 
irregular-tasking and work-sharing pragmas 
to evaluate \cf on a 20-core Intel processor.
We show that \cf significantly improves the 
energy-efficiency with a negligible impact on performance.
We also evaluated \cf by implementing a subset of these benchmarks using
\afp task parallelism~\cite{KVZY2014} to demonstrate that \cf is oblivious
to the parallel programming model. 

In summary, this paper makes the following contributions:

\begin{itemize}

\item \cf, a parallel programming model oblivious C/C++ library for achieving energy efficiency
      in multicore parallel programs running on Intel processors.
\item A novel light-weight runtime for \cf that periodically monitors the
      memory access pattern of a running application and then dynamically
      adapts the core and uncore frequencies using DVFS and UFS, respectively.
\item Evaluation of \cf on a 20-core Intel Xeon Haswell E5-2650 processor by
      using multiple HPC benchmarks and  mini-applications implemented in \omp and \afp programming model. Our evaluation
      shows that \cf can significantly improve energy efficiency with negligible impact on the execution time for several 
      irregular-tasking and work-sharing pragmas.

\end{itemize}

\section{Experimental Methodology}
\label{sec:methodology}

\begin{table*}[]
\centering
\resizebox{0.95\textwidth}{!}{%
\begin{tabular}{|c|c|c|c|c|c|c|c|}
\hline
\multirow{2}{*}{\begin{tabular}[c]{@{}c@{}}Benchmark\\ Name\end{tabular}} & \multirow{2}{*}{Brief Description}                                                        & \multirow{2}{*}{Configuration} & \multirow{2}{*}{\begin{tabular}[c]{@{}c@{}}Parallelism\\ Style\end{tabular}} & \multirow{2}{*}{\begin{tabular}[c]{@{}c@{}}OpenMP\\ Time (sec)\end{tabular}} & \multirow{2}{*}{\begin{tabular}[c]{@{}c@{}}TIPI \\ Range\end{tabular}} & \multicolumn{2}{l|}{Total TIPI Slabs} \\ \cline{7-8} 
                                                                          &                                                                                           &                                &                                                                              &                                                                                          &                                                                        & Distinct          & Frequent          \\ \hline
\uts                                                                       & Unbalanced Tree Search~\cite{OSHJ2006}                                                                    & TIXXL                          & Irregular Tasks                                                              & 69.9                                                                                     & 0-0.004                                                                & 1                 & 1                 \\ \hline
\sorir                                                                       & Successive Over-Relaxation (SOR)~\cite{bull2000benchmark}                                                          & 32Kx32K (200)                  & Irregular Tasks                                                              & 69.1                                                                                     & 0.024-0.028                                                            & 1                 & 1                 \\ \hline
\sorrr                                                                       & \rebuttal{\ditto}                                                                                      & \rebuttal{\ditto}                           & Regular Tasks                                                                & 69.4                                                                                     & 0.024-0.028                                                                   & 1                 & 1                 \\ \hline
\sorws                                                                       & \rebuttal{\ditto}                                                                                      & \rebuttal{\ditto}                           & Work-sharing                                                                 & 68.7                                                                                     & 0.012-0.028                                                                   & 3                 & 1                 \\ \hline
\heatir                                                                      & Heat diffusion (Jacobi-type iteration)~\cite{mitcilkversion}                                                    & 32Kx32K (200)                  & Irregular Tasks                                                              & 76.6                                                                                     & 0.056-0.076                                                            & 4                 & 1                 \\ \hline
\heatrr                                                                      & \rebuttal{\ditto}                                                                                      & \rebuttal{\ditto}                           & Regular Tasks                                                                & 75.5                                                                                     & 0.056-0.072                                                                   & 3                 & 2                 \\ \hline
\heatws                                                                      & \rebuttal{\ditto}                                                                                      & \rebuttal{\ditto}                           & Work-sharing                                                                 & 70.9                                                                                     & 0.012-0.068                                                                   & 11                 & 1                 \\ \hline
\minife                                                                    & Finite Element Mini-Application~\cite{mantevo,heroux2009improving}                                                           & 256x512x512 (200)              & Work-sharing                                                                 & 78.5                                                                                     & 0.068-0.152                                                            & 16                & 1                 \\ \hline
\hpccg                                                                     & \begin{tabular}[c]{@{}c@{}}High Performance Computing \\ Conjugate Gradients~\cite{mantevo,heroux2009improving}\end{tabular} & 256x256x1024 (149)             & Work-sharing                                                                 & 60                                                                                     & 0.060-0.148                                                            & 17                & 1                 \\ \hline
\amg                                                                       & Algebraic Multigrid solver~\cite{amg}                                                                & 256x256x1024 (22)              & Work-sharing                                                                 & 63.7                                                                                     & 0.060-0.332                                                            & 60                & 2                 \\ \hline
\end{tabular}
}
\caption{Description of the benchmarks used in this paper for the evaluation of \cf}
\label{tab:bench}
\end{table*}


\begin{figure}[h]
  \centering
  \includegraphics[width=\linewidth]{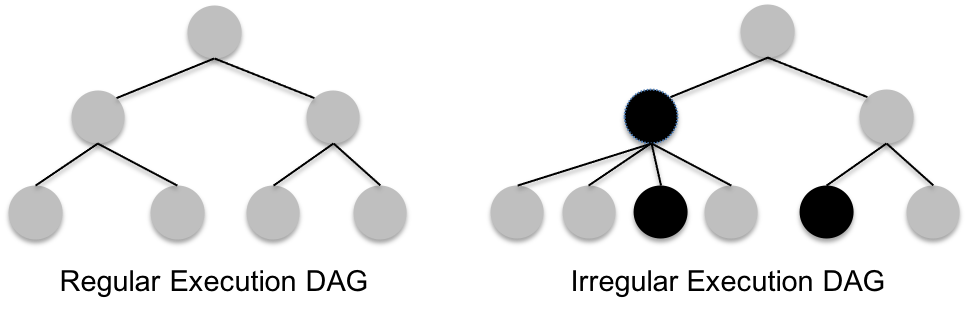}
    \captionof{figure}{\rebuttal{\heatir and \sorir use the irregular execution DAG, whereas \heatrr and \sorrr use the regular execution DAG. Non-root and non-leaf grey and black nodes have degrees three and five, respectively. Each node is parallel task}}
   \label{fig:dag}
\end{figure}

Before presenting the motivating analysis for \cf, we first describe our experimental methodology.
To cover a broad spectrum of parallel applications, we have chosen benchmarks 
based on the following attributes: 
a) memory access patterns, 
b) micro-kernels and real-world mini-applications,
c) execution DAG, 
and d) concurrency decomposition techniques. 
We target six widely-used benchmarks for our experimental evaluation. 
These benchmarks mainly use \omp pragmas, but a few were ported to \afp programming model 
supported by the \hc work-stealing library~\cite{KVZY2014,GMKV2017} 
to evaluate \cf using two different parallel programming models.
We created three variants (both for \omp and \hc) that differ 
in execution DAG and concurrency decomposition technique for two of these 
benchmarks, Heat and SOR.
Two of these variants use dynamic task parallelism
and vary in terms of irregular (\emph{irt}) and regular (\emph{rt}) execution DAGs.
The third variant is a non-recursive implementation that
uses work-sharing (\emph{ws}) based static loop partitioning. 
The technique described in
Chen \etal~\cite{Chen2014} was used for converting the loop-level
parallelism in Heat and SOR into regular and irregular execution DAG, 
as shown in Figure~\ref{fig:dag}. 
In total, we have ten benchmarks/mini-applications. 
Table~\ref{tab:bench} describes them and their respective configurations.

We ran all experiments on an Intel Xeon Haswell E5-2650 v3 20-core processor
with a total of 94GB of RAM. The operating system (OS) was Ubuntu 16.04.7 LTS.
This processor supports core and uncore frequencies between 1.2GHz--2.3GHz
and 1.2GHz--3.0GHz, respectively, both in steps 0.1GHz. 
We used \textjava{MSR-SAFE}~\cite{msrsafe}
for saving and restoring Model-Specific-Register (MSR) values.
The turboboost feature on the processor was disabled, and each benchmark used interleaved
memory allocation policies supported by the \textjava{numactl} library. 
The \hc implementation from the official Github repository with the commit id \textjava{ab310a0} is used.
Both \omp and \hc versions of the benchmarks used the Clang compiler version 3.8.0 with the -O3 flag.
All 20 threads used in the experiments were bound to their respective physical CPUs.
\rebuttal{
Although we use an Intel Haswell processor in our evaluation, more recent Intel processors can use \cf by updating the MSRs specific to them. While the latest AMD processors do support per-core DVFS similar to Intel, based on public documentation available, it is not clear if support for UFS or TIPI measurement exists.}

We evaluated three variants of \cf: a) \cfc that only adapts the CPU frequency,
b) \cfu that only adapts the Uncore frequency, and c) \cf that adapts
both the CPU and Uncore frequencies. We compared these implementations against
each benchmark's \orig execution by setting the Intel power governor
to \textjava{performance} policy. 
The performance power governor fixes the CPU frequency to the maximum.
We chose \textjava{performance} power governor
for \orig execution as this same setting is 
used by several supercomputers in production~\cite{hartuser,cook2017performance}.
\rebuttal{For Cuttlefish-based executions, the power governor is set to userspace to allow 
changes to the CPU frequency through the library.}
\rebuttal{We changed the uncore frequency scaling option to \emph{Auto} in the BIOS,
allowing the Intel firmware to modulate the uncore frequency during the \orig execution.
The algorithm used by Intel is highly sensitive to memory requests.
During \cf executions, the runtime controls the uncore frequency
dynamically inside the library by writing the desired frequency value 
in the UFS MSR (\textjava{0x620}).}
We executed each implementation ten times and reported the mean 
value along with a 95\% confidence interval.  

\section{Motivating Analysis}
\label{sec:motivate}

\begin{figure*}[h]
    \centering
    \subfigure[Variation in JPI during the course of execution]{\label{graph:timelineJPI}\includegraphics[width=0.49\textwidth]{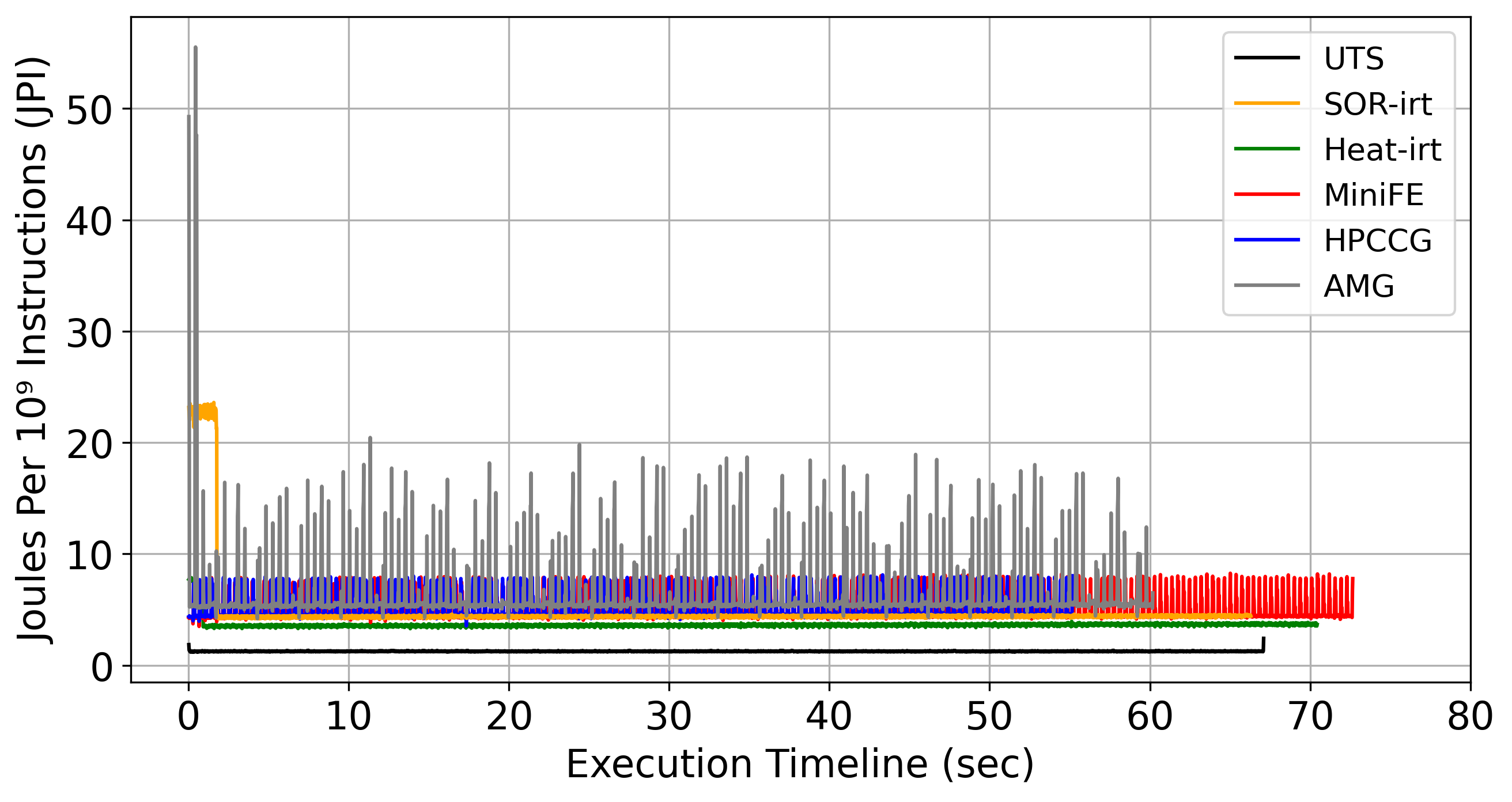}}
    \subfigure[Variation in TIPI during the course of execution]{\label{graph:timelineTIPI}\includegraphics[width=0.49\textwidth]{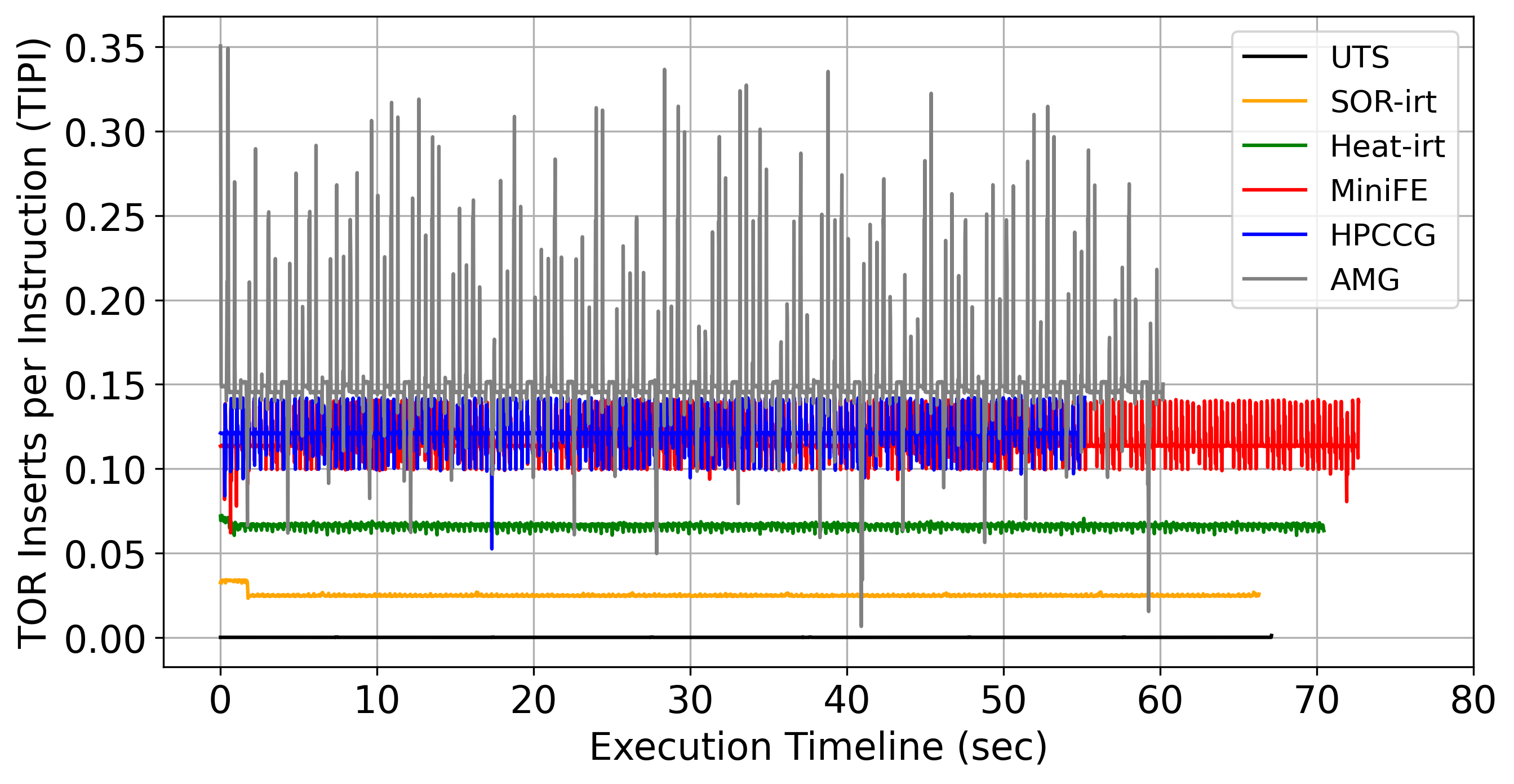}}
   \caption{Relation between TIPI and JPI. For each benchmark, JPI increases with the increase in TIPI}
    \label{graph:motivate1}
\end{figure*}

An application's memory access pattern (MAP) can be primarily classified as
memory-bound or compute-bound. 
Memory-bound applications have a high number of memory accesses
as compared to compute-bound applications.
This section presents two analyses to support
that MAP of an application can be accurately identified 
using MSRs (Section~\ref{sec:tipi}), and changing the
core and uncore frequencies based on MAP can lead to
energy savings (Section~\ref{sec:tipitrend}).

\subsection{TOR Inserts per Instruction (TIPI)}
\label{sec:tipi}

To accurately identify the MAP of an application, we propose
the metric TOR Inserts per Instruction (\emph{TIPI}), which 
is  calculated as the ratio of total TOR Inserts 
(\textjava{TOR_INSERT.MISS_LOCAL + TOR_INSERT.MISS_REMOTE})
and total instructions retired (\textjava{INST_RETIRED.ANY}).
Any requests coming to the LLC from the processor cores are
place in TOR (Table of Request).
\torins MSR~\cite{torinsert} is available
on all Intel processors from Haswell generation and onwards.
This MSR counts the number of memory requests
that come to the socket-specific Last-Level Cache (LLC) from each core.
It supports various Unit Masks (umask) to select the type of memory requests 
to be counted.
\textjava{MISS_LOCAL} counts the misses to the local caches and memory, and
\textjava{MISS_REMOTE} counts the misses to the remote caches and memory as per 
the Intel documentation~\cite{torinsert}.
We consider both these umasks as our experimental machine
is a two-socket NUMA machine (Section~\ref{sec:methodology}). 
The other metric used is Joules per Instructions (\emph{JPI}). 
JPI is calculated as the ratio of the total energy spent
by the processor and the total instructions retired.
The package energy is measured using the Intel RAPL MSRs.

We found that compute-bound and memory-bound applications 
have low and high TIPI values, respectively.
It is also observed that TIPI has a strong correlation with JPI
as an increase in TIPI also increases the JPI. These
observations are reported in Figure~\ref{graph:motivate1}.
The core and uncore frequencies are initially set to the maximum, and 
periodically the TIPI and JPI
for an application is recorded at every \interval intervals.
The \interval is fixed at 20 milliseconds and is empirically derived 
for all experiments reported in this paper. 
The x-axis in Figure~\ref{graph:motivate1} represents the
execution timeline of each benchmark in seconds.
The y-axis in Figure~\ref{graph:timelineTIPI} and Figure~\ref{graph:timelineJPI}
shows the TIPI and JPI, respectively, during the 
benchmark's execution.
The result of only \heatir and \sorir are reported as their variants have similar
behaviours.
As JPI is highly sensitive to TIPI, JPI is used to measure energy efficiency
for a given TIPI. 
In Figure~\ref{graph:timelineJPI}, \sorir has a higher JPI than \heatir, although 
\sorir's TIPI is less than that of \heatir (Figure~\ref{graph:timelineTIPI}). 
For \minife, \hpccg, and \amg, it is observed that TIPI varies throughout the
execution.
However, an increase in TIPI shows an increase in JPI.
This implies that the TIPI v/s JPI trend holds only within an application,
and the same TIPI value could have different JPIs for different applications.
\vspace{-1ex}

\subsection{Analysis of DVFS and UFS with TIPI} 
\label{sec:tipitrend}

\begin{figure*}
    \centering
    \subfigure[Effect of fixing Uncore and changing Core frequency]{\label{graph:mapcore}\includegraphics[width=\textwidth]{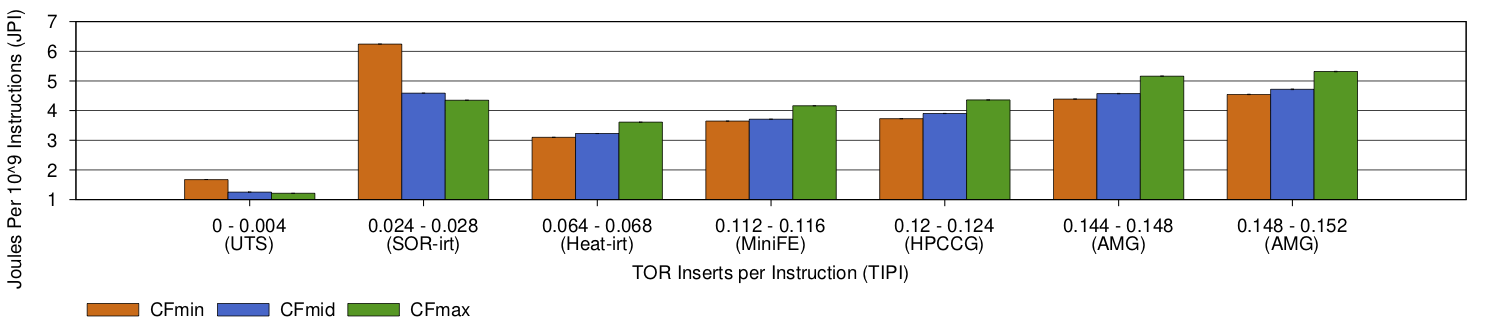}}
    \subfigure[Effect of fixing Core and changing Uncore frequency]{\label{graph:mapuncore}\includegraphics[width=\textwidth]{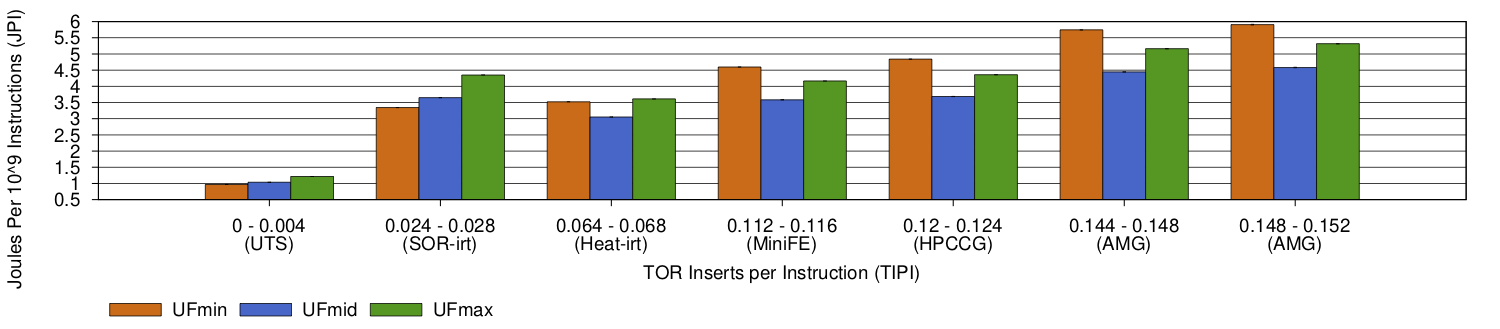}}
   \caption{Motivational analysis to understand the effect of core and uncore frequencies on the JPI at each TIPI}
    \label{graph:motivate2}
\end{figure*}


\cf has an online profiler that explores optimal core and uncore
frequencies for every unique TIPIs discovered during a benchmark 
execution. To limit the search space, we divide each unique TIPIs
in fixed slabs of 0.004 (derived empirically), \ie TIPI values 0.004, 0.005, and 0.007 would be reported
under the TIPI range 0.004-0.008.
Hereafter, a TIPI value is reported in terms of its range instead of the actual value.
Figure~\ref{graph:motivate2} shows the relation between TIPI and JPI for benchmarks
at different core and uncore frequencies.  
In Figure~\ref{graph:mapcore}, the Uncore Frequency (UF) is set to max (3.0 GHz) 
and each benchmark is executed with three different Core Frequencies (CF) -- 
min (1.2 GHz), mid (1.8 GHz) and max (2.3 GHz).
In Figure~\ref{graph:mapuncore}, the CF is set to max (2.3 GHz) 
and each benchmark is executed with three different UFs -- 
min (1.2 GHz), mid (2.1 GHz) and max (3.0 GHz).
In each execution of a benchmark, we record the TIPI and JPI at every \interval
and then calculate the average JPI for the frequently occurring TIPIs.
TIPIs found in more than 10\% of total \interval samplings
are mentioned as frequently occurring TIPIs during an execution.

\rebuttal{We can observe that \uts and \sorir have
a low TIPI range and are compute-bound
as their JPI decreases with increasing CF and JPI increases with increasing UF. 
\heatir, \minife, \hpccg, and \amg have a high TIPI range
and are memory-bound applications as they behave precisely opposite
to \uts and \sorir. JPI of these four benchmarks
increases with increasing the CF, and their JPI
decreases with increasing the UF.
}
Although \heatir, \minife, \hpccg, and \amg are memory-bound,
max uncore frequency is not apt for their TIPI range.
This analysis implies that when TIPI is low,
CPU cores should run at a higher frequency,
and uncore should run at a lower frequency.
In contrast, when TIPI is high, uncore should run at a higher frequency, and CPU cores
should run at a lower frequency as the
latter would frequently halt due to the memory access latency.

\section{Design and Implementation}
\label{sec:impl}



The previous section highlighted that TIPI could be used 
to distinguish the MAP of an application and for estimating
its optimal core and uncore frequencies.
Hence, to achieve energy efficiency in an 
application with minimal loss in performance,
it is vital to use a low-overhead online profiler 
to discover the TIPI ranges dynamically and appropriately set 
the core and uncore frequencies. We approach the problem 
using a lightweight daemon thread, \cf, that runs in tandem with the application to
monitor the TIPI and JPI by activating itself after every \interval durations to
minimize the time-sharing of the CPU with the application. 
In their application, programmers only need to use two API \rebuttal{functions}, \cfstart and \cfstop, to
define the scope requiring an energy-efficient execution.
Whenever \cf discovers a
new TIPI range, it uses DVFS to explore the optimal core frequency,
followed by UFS to explore the optimal uncore frequency for this
TIPI range. An optimal frequency is the one that has the lowest JPI.
\cf start its execution without any prior information
on TIPI ranges and optimal frequencies. The insight is to minimize the application slowdown
due to frequency exploration by compacting the exploration
range on the fly based on the optimal core and uncore frequencies
of other TIPI range discovered since the start of this application. 


\subsection{\cf daemon loop}
\label{sec:cfinit}

\begin{algorithm}
\begin{algocolor}
\DontPrintSemicolon

\SetAlgoLined
    \cfa{prev} $\gets$ \cfa{max}$; $ \ufa{prev} $\gets$ \ufa{max}\;
    set\_freq(\cfa{prev}, \ufa{prev})\; 						\label{loop:ref1}
    sleep(warmup duration)\;								\label{loop:ref7}
  \cfa{next} $\gets$ \ufa{next}  $\gets$ -1 \;
  \ncurr $\gets$ \nprev $\gets$ NULL \tcp*{TIPI LinkedList nodes}					
  \While{ shutdown not initiated from \textjava{stop} API}{
  Read TIPI and JPI values from MSR\;                                                   \label{loop:ref3}
    \uIf{TIPI NOT found in LinkedList}{							\label{loop:ref4}
        \ncurr $\gets$ insert\_in\_sortedLinkedList(TIPI)\;
        \cfa{LB} $\gets$ \cfa{min}$; $ \cfa{UB} $\gets$ \cfa{max} \tcp*{for \ncurr}	\label{loop:ref8}
        \ufa{LB} $\gets$ \ufa{min}$; $ \ufa{UB} $\gets$ \ufa{max} \tcp*{for \ncurr}	\label{loop:ref9}
	\tcp{Section~\ref{sec:multitipi1}}
        Reduce \cfa{LB}$/$\cfa{RB} if LinkedList\textsubscript{size}>1\;		\label{loop:ref18}
	\cfa{next} $\gets$ find(CF, JPI, \cfa{prev}, \nprev, \ncurr)\;
	\ufa{next} $\gets$ \ufa{max}\;
    }
    \Else{
        \ncurr $\gets$ fetch\_from\_sortedLinkedList(TIPI)\;
	\uIf{\ncurr.\cfa{opt}$\And$\ncurr.\ufa{opt} = -1 }{
        	\cfa{next} $\gets$ find(CF, JPI, \cfa{prev}, \nprev, \ncurr)\;	
		\ufa{next} $\gets$ \ufa{max}\;
        	\If{\ncurr.\cfa{next} = \ncurr.\cfa{opt}}{				\label{loop:ref14}
           		Estimate \ufa{LB}$\And$\ufa{RB} (Algorithm~\ref{algo:guess})\;	\label{loop:ref15}
	   		\tcp{Section~\ref{sec:multitipi1}}
                        Reduce \ufa{LB}$/$\ufa{RB} if LinkedList\textsubscript{size}>1\; \label{loop:ref19}
			\ufa{next} $\gets$ \ncurr.\ufa{RB}\;
        	}
    	}
    	\uElseIf{\ncurr.\ufa{opt} = -1 }{
		\cfa{next} $\gets$ \ncurr.\cfa{opt}\;
        	\ufa{next}$\gets$find(UF, JPI, \ufa{prev}, \nprev, \ncurr)\;
    	}
    	\Else{
		\cfa{next} $\gets$ \ncurr.\cfa{opt}\; 
                \ufa{next} $\gets$ \ncurr.\ufa{opt}\;
    	}
    }
    
    set\_freq(\cfa{next} , \ufa{next} )\;						\label{loop:ref5}	
    \nprev $\gets$ \ncurr\;
    \cfa{prev} $\gets$ \cfa{next}$;$ \ufa{prev} $\gets$ \ufa{next}$;$\;
    
  sleep ( \interval )\;									\label{loop:ref6}
 }
\caption{\rebuttal{Cuttlefish daemon thread method}}
\label{algo:policy}
\end{algocolor}
\end{algorithm}

\cf daemon thread is spawned by the API \cfstart and is pinned to a fixed
core. \rebuttal{Algorithm~\ref{algo:policy} lists the pseudocode
implementation of this daemon thread}. It sets the core and uncore frequencies
to the maximum (\rebuttal{Line~\ref{loop:ref1}}). It runs in a loop where it first calculates the TIPI and 
JPI for the \rebuttal{whole processor (Line~\ref{loop:ref3}}), followed by 
execution of the \cf runtime policy (\rebuttal{Line~\ref{loop:ref4}--Line~\ref{loop:ref5}}), and finally 
goes back to sleep for \interval duration (\rebuttal{Line~\ref{loop:ref6}}). 
The implementation for measuring the TIPI and JPI in \cf is inspired by the RCRtool~\cite{porterfield2010RCR}.
It continues the execution of this loop until \cfstop is called inside the user application.
From Figure~\ref{graph:timelineTIPI} and Figure~\ref{graph:timelineJPI},
we can observe that TIPI and JPI fluctuate heavily at the beginning of 
the execution timeline. This fluctuation is prominent across all three variants of Heat and SOR
and \amg (seven out of ten benchmarks), whereas it's minuscule
in the remaining three benchmarks. This instability is due to cold caches
at the start of execution, but it becomes stable after a while.
Hence, to avoid recording unstable values of TIPI and JPI, the \cf
daemon loop activates only after a warmup duration of two seconds (\rebuttal{Line~\ref{loop:ref7}}).

\subsection{\cf runtime policy}
\label{sec:cfpolicy}

After every \interval, the \cf policy (\rebuttal{Line~\ref{loop:ref4}--Line~\ref{loop:ref5}}) 
is invoked where it uses DVFS and UFS to determine Optimal Core Frequency (\ocf)
and Optimal Uncore Frequency (\ouf) for a given TIPI range.
\cf maintains a sorted doubly linked list of unique TIPI ranges
discovered during an execution. This list is empty 
at the beginning (\rebuttal{Line~\ref{loop:ref4}}). 
Each node in this linked list has the following fields: 
TIPI range,
arrays to store JPI for each core and uncore frequencies,
latest exploration range for core and uncore frequencies,
\ocf, and \ouf. Figure~\ref{fig:explore}(a) shows one such node
in this list for a hypothetical processor having seven frequency levels,
\textjava{A-G}, for both core and uncore. \textjava{A}
and \textjava{G} are the lowest and highest frequencies, respectively, in
this processor. 
To explain the \cf runtime policy's working, we use this 
same hypothetical processor in all our discussions hereafter. The \rebuttal{pseudocode
implementation of core and uncore frequency exploration is shown in Algorithm~\ref{algo:find}.}

\begin{algorithm}
\begin{algocolor}
\DontPrintSemicolon

\SetAlgoLined
  \KwIn{type, \jpi{curr}, \fl{curr}, \nprev, \ncurr}
  \KwOut{\fl{next}}
    ptr = \ncurr.FQ\_table$[$type$]$ \tcp{type is CF/UF}
    
    \If{ptr.\fl{LB} $\And$ ptr.\fl{RB} are adjacent} {					\label{find:ref16}
        Choose \fl{opt} from \fl{LB} $\And$ \fl{RB} \tcp{Figure~\ref{fig:impl2}}
	\textbf{return} ptr.\fl{opt}\;							
    }											\label{find:ref17}
    \tcp{Discard JPI in TIPI transition phase}					
    \If{\nprev = \ncurr}{								\label{find:ref20}
    	\ncurr.JPI\_table$[$type$][$\fl{curr}$]$ $\gets$ average(\jpi{curr})\; 	\label{find:ref10}
    }											\label{find:ref21}
    \tcc{\jpi{avg} at any FQ is average of 10 readings.
    Hence, Line~\ref{exp:if3} / Line~\ref{exp:elif1} equates to true during TIPI transition due to incomplete \jpi{avg}}
    \uIf{\jpi{avg} NOT exists for ptr.\fl{RB}}{			\label{exp:if3}
	\textbf{return} ptr.\fl{RB}\;
    }
    \ElseIf{\jpi{avg} NOT exists for ptr.\fl{RB-2}}{		\label{exp:elif1}
        \textbf{return} ptr.\fl{RB-2}\;
    }
    \uIf{\jpi{avg} at ptr.\fl{RB-2} is less than ptr.\fl{RB}}{				\label{find:ref11}
        ptr.\fl{RB} $\gets$ ptr.\fl{RB-2}\;
	\fl{next} = (ptr.\fl{RB} - ptr.\fl{LB} > 2) ? ptr.\fl{RB-2} : ptr.\fl{LB}\;	\label{find:ref12}
    }
    \Else{
        \fl{next} $\gets$ ptr.\fl{LB} $\gets$ ptr.\fl{RB-1}\;
    }
    \If{ptr.\fl{LB} $\And$ ptr.\fl{RB} are same}{
	\fl{next} $\gets$ ptr.\fl{opt} $\gets$ ptr.\fl{RB}\;				\label{find:ref13}
    }
    \tcp{Section~\ref{sec:multitipi2}}
    Update \fl{LB} or \fl{RB} of other TIPIs if LinkedList\textsubscript{size}>1\;	\label{find:ref22}
    \textbf{return} \fl{next}\;  
\caption{\rebuttal{Method \textjava{find} for CF / UF exploration}}
\label{algo:find}
\end{algocolor}
\end{algorithm}

\subsection{Frequency exploration in single TIPI-range}
\label{sec:singletipi}


\begin{figure}[h]
  \centering
  \includegraphics[width=\linewidth]{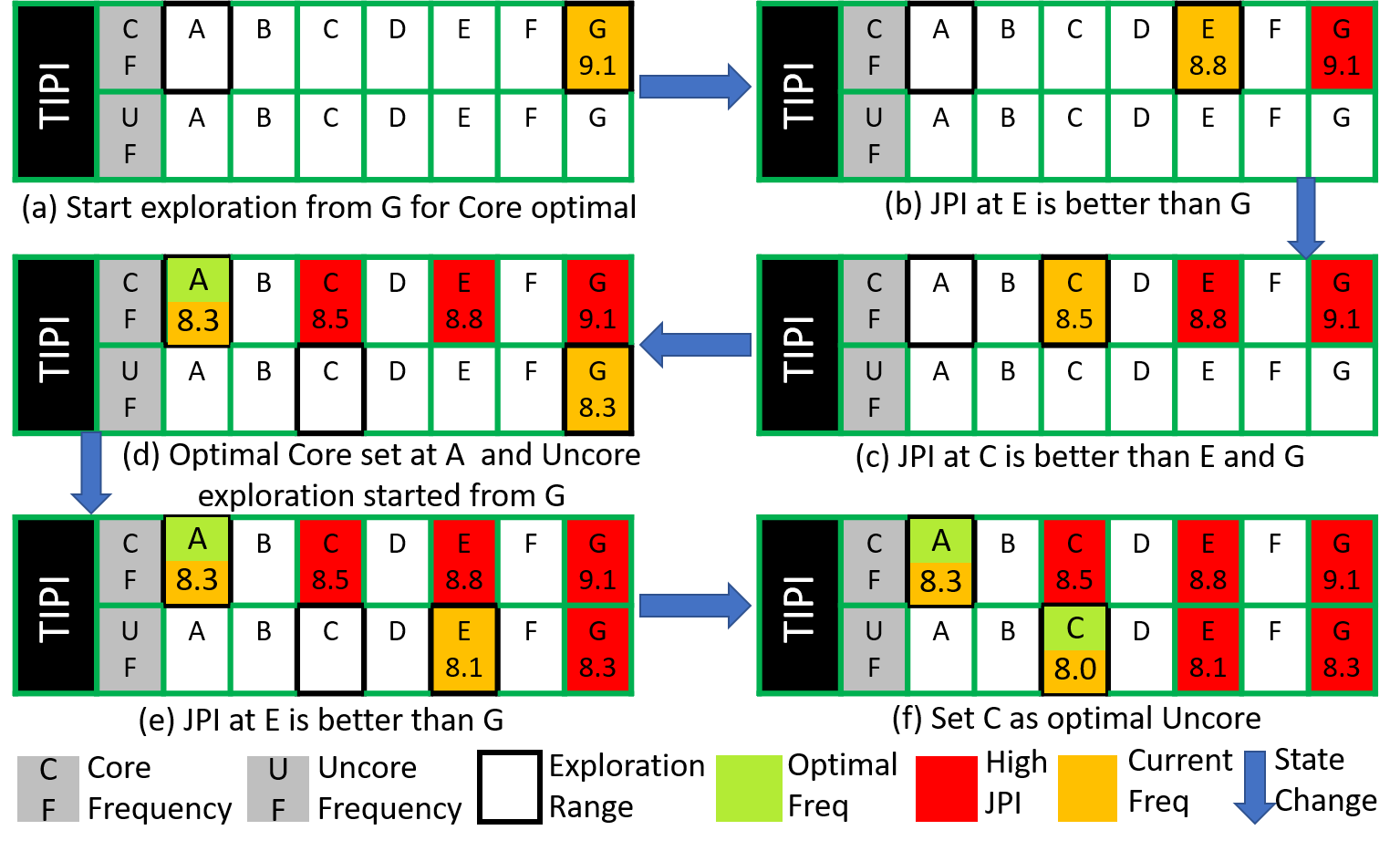}
    \captionof{figure}{Frequency exploration for a single TIPI}
   \label{fig:explore}
\end{figure}

Here, we explain the steps followed by the \cf in
exploring the \ocf and \ouf for benchmarks having a single
TIPI range (Table~\ref{tab:bench} lists the total number of 
distinct TIPI slabs in each benchmark). 
\rebuttal{\cf starts the exploration first for the \ocf within the default exploration
range with the left bound core and uncore frequencies (\clb and \ulb, respectively) set to minimum, 
and the right bound core and uncore frequencies (\crb and \urb, respectively) 
set to maximum (Algorithm~\ref{algo:policy}, Line~\ref{loop:ref8}--Line~\ref{loop:ref9}).} 
For the hypothetical processor,
this exploration is shown in Figure~\ref{fig:explore}, where \clb=\cfl{A} and \crb=\cfl{G}.
As execution starts with the frequencies set to \cfl{G} and \ufl{G}, 
after the first \interval, the JPI value for \cfl{G} is recorded.
To ensure stability in the JPI value, \cf computes the average JPI at \cfl{G} across ten
\interval. Henceforth, the JPI value reported for any \textjava{CF} and \textjava{UF}
in \cf is an average of ten readings \rebuttal{(Algorithm~\ref{algo:find}, Line~\ref{find:ref10}).
Frequency exploration cannot continue until an average of ten readings of JPI is 
available at a given frequency.} 
Once the JPI is finalized for \cfl{G} (Figure~\ref{fig:explore}(a)), \cf will use DVFS
to set the core frequency at \cfl{E} without changing the uncore 
\rebuttal{(Algorithm~\ref{algo:policy}, Line~\ref{loop:ref5})}.
Frequencies are always explored at steps of two to minimize the exploration steps.
\cf will then compare the JPIs at \cfl{G} and \cfl{E} (Figure~\ref{fig:explore}(b)).
\rebuttal{This comparison is performed at
Line~\ref{find:ref11}--Line~\ref{find:ref12} in Algorithm~\ref{algo:find}.}
As \cfl{E} has lower JPI than \cfl{G}, 
\cf will update \crb=\cfl{E}
and then set the frequency at \cfl{C}.
As it turns out that \cfl{C} has lower JPI than \cfl{E}, now \crb=\cfl{C} 
(Figure~\ref{fig:explore}(c)), and the frequency would be set to \cfl{A}.
As even \cfl{A} has lower JPI than \cfl{C}, \clb=\crb=\cfl{A} \rebuttal{(Algorithm~\ref{algo:find}, Line~\ref{find:ref13})}, 
thereby making it as \ocf for this TIPI (Figure~\ref{fig:explore}(d)). 

\rebuttal{
\cf explores both core and uncore frequencies linearly at the steps of 
two instead of using binary search for reducing
the total number of explorations and performance degradation. 
Except when the optimal frequency lies at the boundary of the exploration range,
\cf cannot use the naive binary search algorithm. 
When the optimal frequency lies between the exploration range, JPI will
increase when moving left/right from the optimal frequency. For example, 
assume that JPI\textsubscript{A}<JPI\textsubscript{G}
and \textjava{E} is the optimal frequency for some TIPI (Figure~\ref{fig:explore}).
Applying a binary search to find \textjava{E} would require measuring
JPIs at \textjava{mid}, \textjava{mid+1}, and \textjava{mid-1} at each
split. This would require more explorations as compared
to the linear search (steps of two). For the worst-case scenario (optimal at default minimum),
the total number of explorations for \ocf on our Intel Haswell processor using linear 
search would be six (\textjava{total_frequencies/2}) compared to eight
by using the modified binary search. Also, as the JPI at each frequency is an average
of ten values, exploring \textjava{mid}, \textjava{mid+1}, and \textjava{mid-1} frequencies in
binary search exploration would cause more performance degradation
than linear search (highest to lowest). 
}

After exploring \ocf=\cfl{A}, \cf will start the exploration for \ouf by fixing the core frequency to \cfl{A} \rebuttal{(Algorithm~\ref{algo:policy}, Line~\ref{loop:ref14})}. 
However, for \ouf, \cf would \emph{not} explore the default exploration range of \ulb=\ufl{A} to \urb=\ufl{G}.
Instead, it uses Algorithm~\ref{algo:guess} to estimate the uncore exploration range \rebuttal{(Algorithm~\ref{algo:policy}, Line~\ref{loop:ref15})}.
Algorithm~\ref{algo:guess} is based on our observation in Section~\ref{sec:tipitrend}
that a \emph{high} core frequency as optimal implies a \emph{low} uncore frequency as optimal, and vice-versa.
Hence, \ocf=\cfl{max} implies that \ouf=\ufl{min}, and \ocf=\cfl{min}
implies \ouf=\ufl{max}.
Now, for estimating the optimal uncore (\oufest) based on the \ocf,
we use the curve-fitting technique, where we map the coordinates (\cfl{max}, \ufl{min})
and (\cfl{min}, \ufl{max}) on a straight line. It is shown in Lines 2-3
of Algorithm~\ref{algo:guess}.
We also saw the latest Intel processor's trend that the number of uncore frequencies and core frequencies is roughly similar. 
Hence, to prepare a short exploration range for \textjava{UF}, 
we calculate the ratio of the number of uncore frequencies and core frequencies.
To get a modest exploration space, we are multiplying the ratio by a constant of 4.
This ratio calculation is shown in Line~\ref{ufest:ref1} of Algorithm~\ref{algo:guess}.
Line~\ref{ufest:ref2}--Line~\ref{ufest:ref3} of Algorithm~\ref{algo:guess} shows the \ulb and \urb 
calculation that uses the range and ratio described above.
By using this Algorithm, \cf uses the exploration range of \ulb=\ufl{C} and \urb=\ufl{G},
and continues the \ouf exploration similar to the \ocf exploration 
(Figure~\ref{fig:explore}(e) and Figure~\ref{fig:explore}(f)).


\begin{algorithm}
\DontPrintSemicolon
\SetAlgoLined
  \KwIn{\cfa{opt}}
  \KwOut{\ufa{LB}, \ufa{RB}}
  \range $\gets$ 4 * (\ufa{max} - \ufa{min} + 1)/(\cfa{max} - \cfa{min} + 1)\; \label{ufest:ref1}
  $\ratio$ $\gets$ (\ufa{max} - \ufa{min})/(\cfa{max} - \cfa{min})\;
  \ufa{opt\_est} $\gets$ \ufa{max} - ($\ratio$ * (\cfa{opt} - \cfa{min}))\;
  \ufa{LB} $\gets$ max(\ufa{min} , \ufa{opt\_est} - \range/2)\;			\label{ufest:ref2}
  \ufa{RB} $\gets$ min(\ufa{max} , \ufa{opt\_est} + \range/2)\;
    \If{\ufa{max} - \ufa{opt\_est} $\leq$ \range/2}
    {
        \ufa{LB} $\gets$ \ufa{LB} - (\ufa{opt\_est} + \range/2 - \ufa{max})\;
    }
    \If{\ufa{opt\_est} - \ufa{min} $\leq$ \range/2}
    {
        \ufa{RB} $\gets$ \ufa{RB} + (\ufa{min} - (\ufa{opt\_est} - \range/2))\;
    }											\label{ufest:ref3}

\caption{Algorithm to find \textjava{UF} exploration range}
\label{algo:guess}
\end{algorithm}

\begin{figure}[h]
  \centering
  \includegraphics[width=\linewidth]{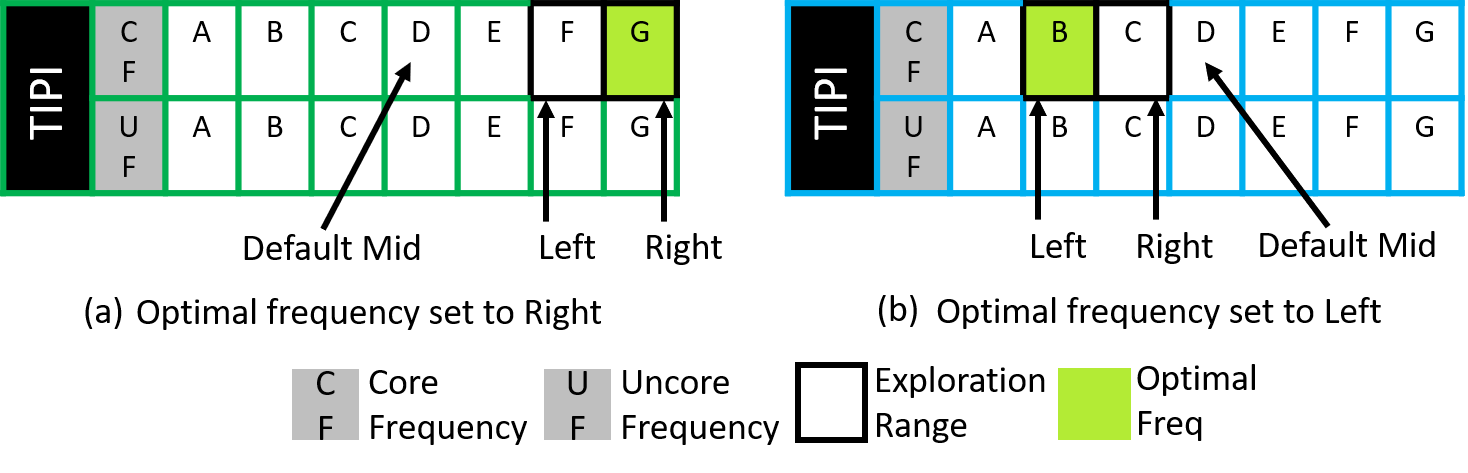}
    \captionof{figure}{Determining \ocf when \crb-\clb=1}
   \label{fig:impl2}
\end{figure}

Figure~\ref{fig:explore} depicted the situation where reducing the frequency 
was reducing the JPI. However, there will be scenarios where
reducing the frequency increases the JPI. These situations are 
depicted in Figure~\ref{fig:impl2} when low bound and high bound
frequencies are adjacents \rebuttal{(Algorithm~\ref{algo:find}, Line~\ref{find:ref16}--Line~\ref{find:ref17})}. 
In Figure~\ref{fig:impl2}(a),
JPI at \cfl{E} was higher than that at \cfl{G}. In this
case, \clb=\cfl{F}. However, as the \clb
and \crb (\cfl{G}) are consecutive frequencies,
\ocf will be set to \cfl{G} to minimize the loss in
performance as high \textjava{CF} indicates a compute-bound MAP. 
Figure~\ref{fig:impl2}(b) depicts a situation
where JPI at \cfl{A} was higher than that at \cfl{C}.
In this case, \clb=\cfl{B}. Again, left bound \clb
and \crb (\cfl{C}) are consecutive frequencies, but
this time \ocf will be set to \cfl{B} to maximize the energy efficiency 
as low \textjava{CF} indicates a memory-bound MAP.
These two scenarios depicted for \textjava{CF}
can also occur for \textjava{UF}. 


\subsection{Frequency exploration for subsequent TIPIs}
\label{sec:multitipi1}


\begin{figure}[h]
  \centering
  \includegraphics[width=\linewidth]{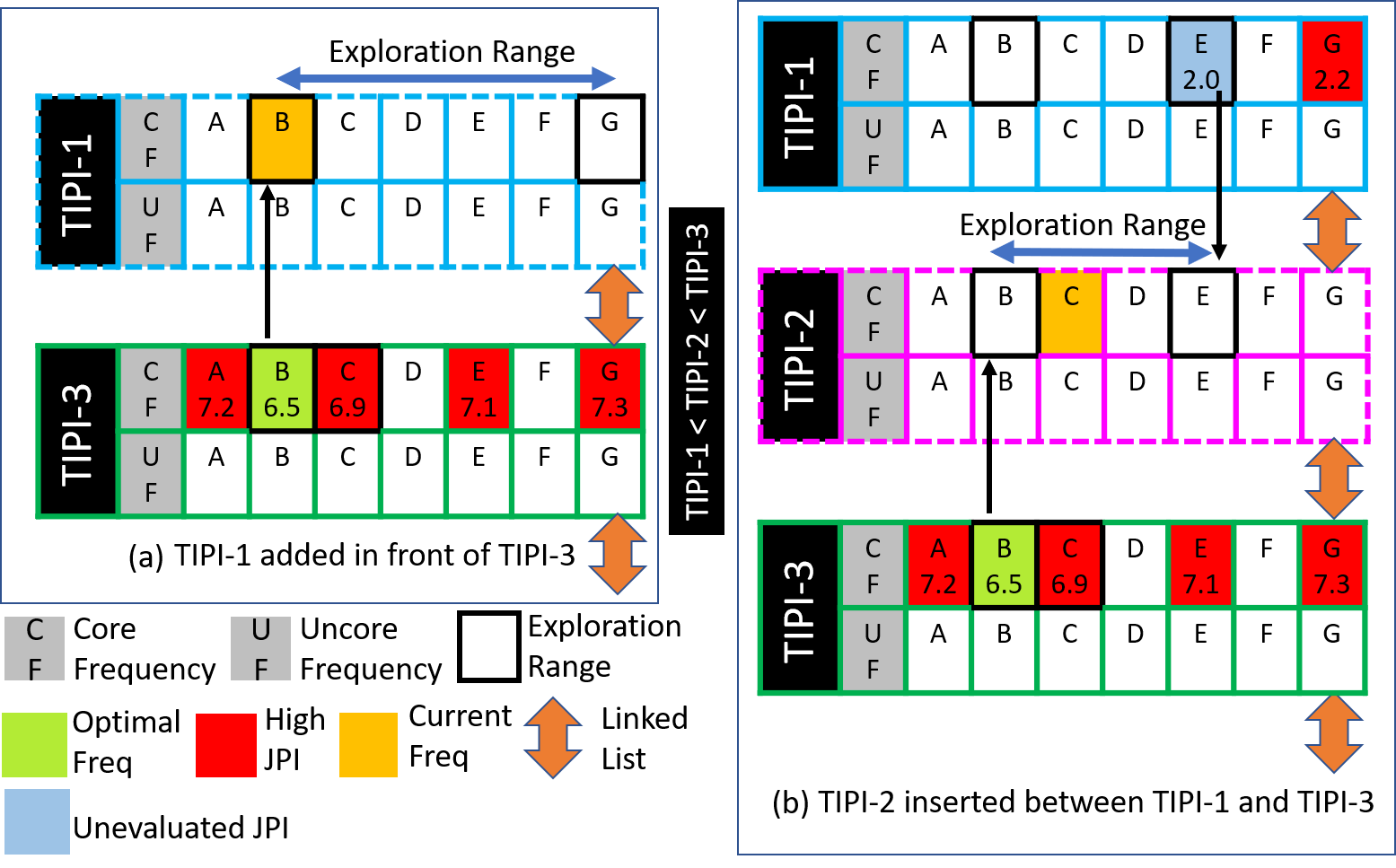}
    \captionof{figure}{Insertion of a node for a newly found TIPI in the doubly linked list, and determining the \textjava{CF} exploration range for this new TIPI}
   \label{fig:impl3}
\end{figure}

\begin{figure}[h]
  \centering
  \includegraphics[width=\linewidth]{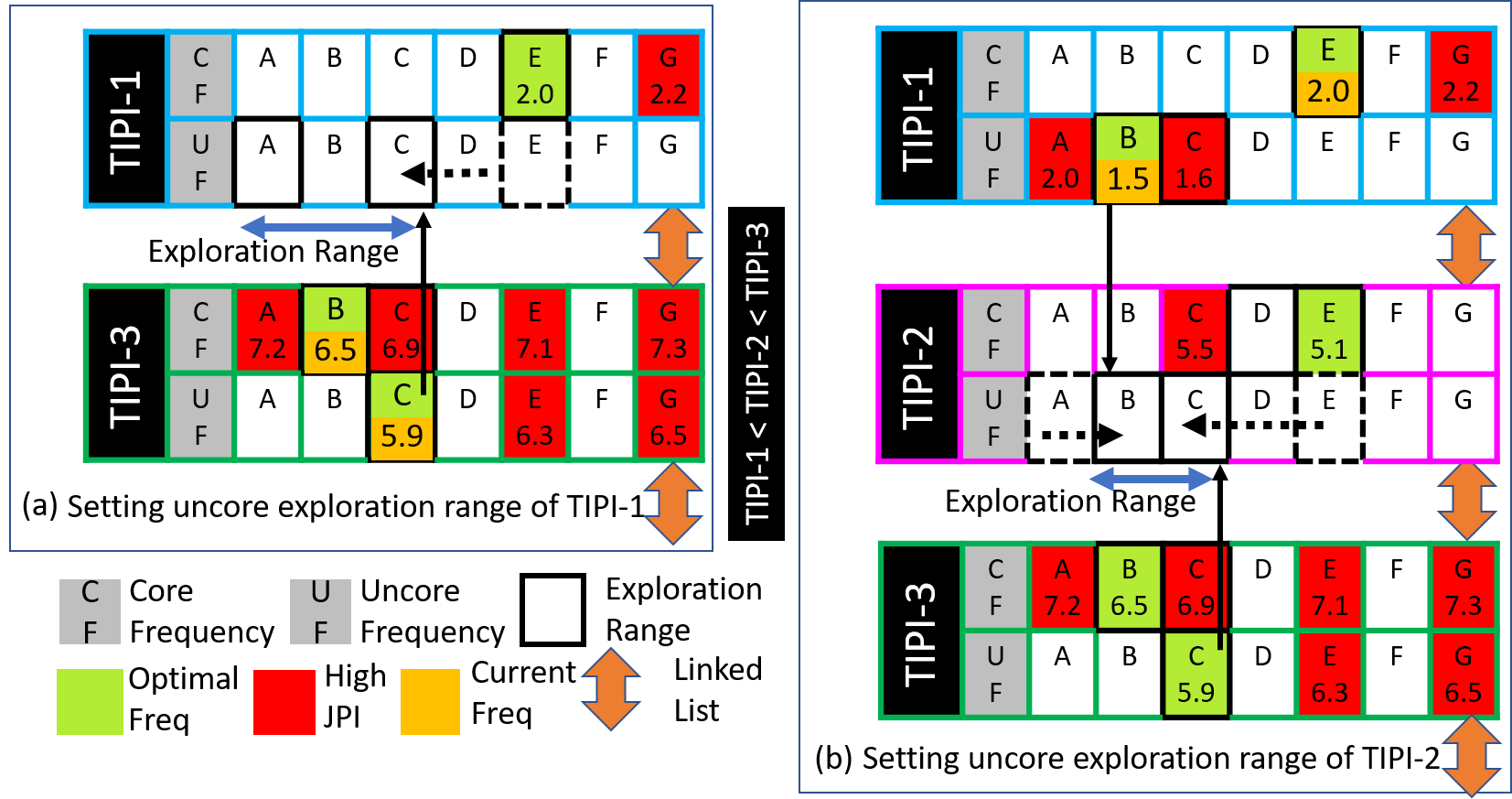}
    \captionof{figure}{Determining the \textjava{UF} exploration range based the \ouf of adjacent TIPI nodes in doubly linked list}
   \label{fig:impl5}
\end{figure}

An application could have a variety of MAP, \eg in \amg, we found 60 distinct
MAPs (Table~\ref{tab:bench}). Using the default \textjava{CF} exploration range 
\clb=\cfl{A} and \crb=\cfl{G} for every newly discovered TIPIs
can easily degrade the application performance, especially on a processor
supporting a wide range of frequencies. Hence, \cf uses the default \clb and \crb
only for the first TIPI \rebuttal{(Algorithm~\ref{algo:policy}, Line~\ref{loop:ref8}--Line~\ref{loop:ref9})}. 
For subsequent TIPIs, it sets \clb, and \crb
based on the \ocf and the latest values of \clb and \crb in previously 
discovered TIPI ranges.
Here, the insight is to use a sorted doubly linked list of
TIPI ranges, where moving from left to right in the linked list signifies a shift from compute-bound
MAP to memory-bound MAP. Hence, when a new TIPI node is inserted in the 
linked list, \cf can look up the \ocf, \clb and \crb of its adjacent nodes (left and right)
to dynamically decide the \clb and \crb for this new TIPI node.
Compared to \textjava{CF}, the exploration range
of \textjava{UF} is already smaller (Algorithm~\ref{algo:guess}). Still, \cf 
attempts to reduce this exploration range even further for subsequent TIPIs by
following the same insight mentioned above for \textjava{CF}.
This optimization is shown in Figure~\ref{fig:impl3} for \textjava{CF} exploration \rebuttal{(Algorithm~\ref{algo:policy}, Line~\ref{loop:ref18})}
and in Figure~\ref{fig:impl5} for \textjava{UF} exploration \rebuttal{(Algorithm~\ref{algo:policy}, Line~\ref{loop:ref19})}.

Figure~\ref{fig:impl3}(a) shows an execution phase \textjava{N} 
(time elapsed since \cf started is \textjava{N}$\times$\interval),
at which MAP was pointing to TIPI-3, but a new TIPI range, TIPI-1,
was discovered by the \cf. Comparing TIPI-1 with other TIPIs in the 
linked list, it is found that TIPI-1 should be added in the front.
The position of the TIPI-1 signifies that it is compute-bound \emph{relative} to the
TIPI-3, \ie \ocf for TIPI-1 will be the same or greater than the \ocf for TIPI-3. 
Hence, \clb=\cfl{B} and \crb=\cfl{G} for TIPI-1. As there was a TIPI transition
in the last \interval phase, JPI calculated for \cfl{B} is not recorded in the \textjava{CF}
table for TIPI-1 \rebuttal{(Algorithm~\ref{algo:find}, Line~\ref{find:ref20}--Line~\ref{find:ref21})}. 
It will be recorded after the next \interval phase.
After a few execution duration with the MAP still pointing to TIPI-1, 
another new TIPI, TIPI-2, was discovered by the \cf.
This is shown in Figure~\ref{fig:impl3}(b).
At this time, \cf calculated the JPI for \cfl{C}
to compare it with the JPI at \cfl{E} in TIPI-1.
However, it won't be able to do so because the new TIPI-2
was discovered in the last \interval. By comparing TIPI-2 with other TIPIs, it found that
the TIPI-2 node should be inserted between the TIPI-1 and 
TIPI-3 nodes in the linked list. This signifies that TIPI-2 
is memory-bound relative to the TIPI-1 but compute-bound relative
to the TIPI-3. Hence, going by the same algorithm as in TIPI-1,
TIPI-2's \clb and \crb should be the \ocf for TIPI-3 and TIPI-1, respectively.
However, as the \ocf for TIPI-1 is still not found, TIPI-2's \crb
will be set to \crb of TIPI-1 (\cfl{E}). Due to the TIPI transition in the
last phase, the JPI (\cfl{C}) of the last \interval will not be recorded in TIPI-2.

Figure~\ref{fig:impl5} demonstrates the same optimization discussed above,
but only for uncore exploration.
Figure~\ref{fig:impl5}(a) shows an execution phase when \cf
has found the \ocf for TIPI-1, and now it has to start the exploration of \ouf for TIPI-1.
By using Algorithm~\ref{algo:guess}, \ulb=\ufl{A} and \urb=\ufl{E} for TIPI-1. 
As TIPI-1 is compute-bound
relative to the TIPI-3, its \ouf will be the same or lower than 
the \ouf (\ufl{C}) for TIPI-3. Hence, \urb=\ufl{C} for TIPI-1.
This is precisely opposite to what was done in the case of \textjava{CF}. 
Figure~\ref{fig:impl5}(b) shows an execution phase when \cf
has found the \ocf for TIPI-2, and now it has to start 
exploring \ouf for TIPI-2. It sets the \ouf of TIPI-3
as \urb for TIPI-2 and \ouf of TIPI-1 as \ulb of TIPI-2.
TIPI-2 is memory-bound relative to TIPI-1. Hence its
\ouf will be the same or higher than the \ouf of TIPI-1.


\subsection{Revalidating frequency exploration range}
\label{sec:multitipi2}

Previous sections described the optimizations carried out in \cf
to reduce the frequency exploration range when the 
exploration is about to start. This section describes
the third and final optimization in \cf to
shrink the exploration range even further after the exploration
has begun \rebuttal{(Algorithm~\ref{algo:find}, Line~\ref{find:ref22})}.

\begin{figure}[h]
    \centering
    \includegraphics[width=\linewidth]{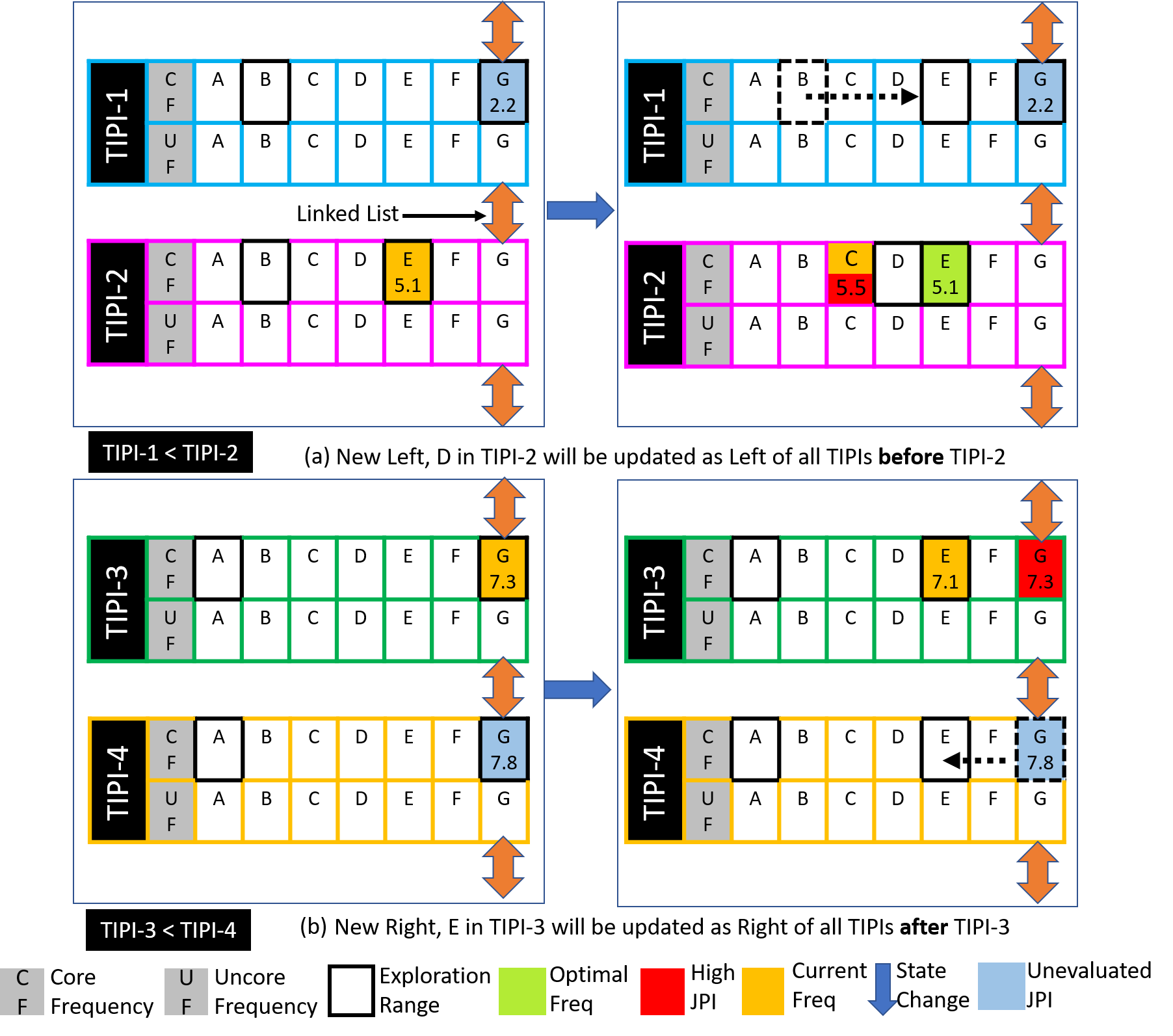}
    \captionof{figure}{Revalidation of \textjava{CF} exploration range}
   \label{fig:impl4}
\end{figure}

Figure~\ref{fig:impl4}(a) shows a phase change from left to right
after some \interval. There are two TIPIs, TIPI-1 and TIPI-2, in the
linked list, where TIPI-1 is compute-bound relative to TIPI-2.
Before the phase change, MAP was set to TIPI-2 with its 
\clb=\cfl{B} and \crb=\cfl{E}. JPI for \cfl{E} has been
evaluated for TIPI-2. Now \cf has to calculate the JPI for \cfl{C}
for TIPI-2. TIPI-1's \clb=\cfl{B} and \crb=\cfl{G}. After some
\interval, \cf evaluates the JPI for \cfl{C} and found it
more than the JPI for \cfl{E}. Hence, for 
TIPI-2 (Figure~\ref{fig:impl4}(a)), it would update the
\clb=\cfl{D} and set \ocf=\cfl{E} as \clb and \crb are 
consecutive frequencies (Section~\ref{sec:singletipi}).
However, as TIPI-1 is compute-bound relative to TIPI-2,
\cf will also update the \clb=\cfl{E} (\ocf for TIPI-2).
In Figure~\ref{fig:impl4}(b), after the phase change,
\cf calculates that JPI at \cfl{E} is better than that at
\cfl{G} for TIPI-3. Hence, it updates the \crb=\cfl{E}
for TIPI-3. However, as TIPI-4 is memory-bound relative
to the TIPI-3, \cf will also update the \crb of TIPI-4
to that of TIPI-3. 




\begin{figure}[h]
  \centering
  \includegraphics[width=\linewidth]{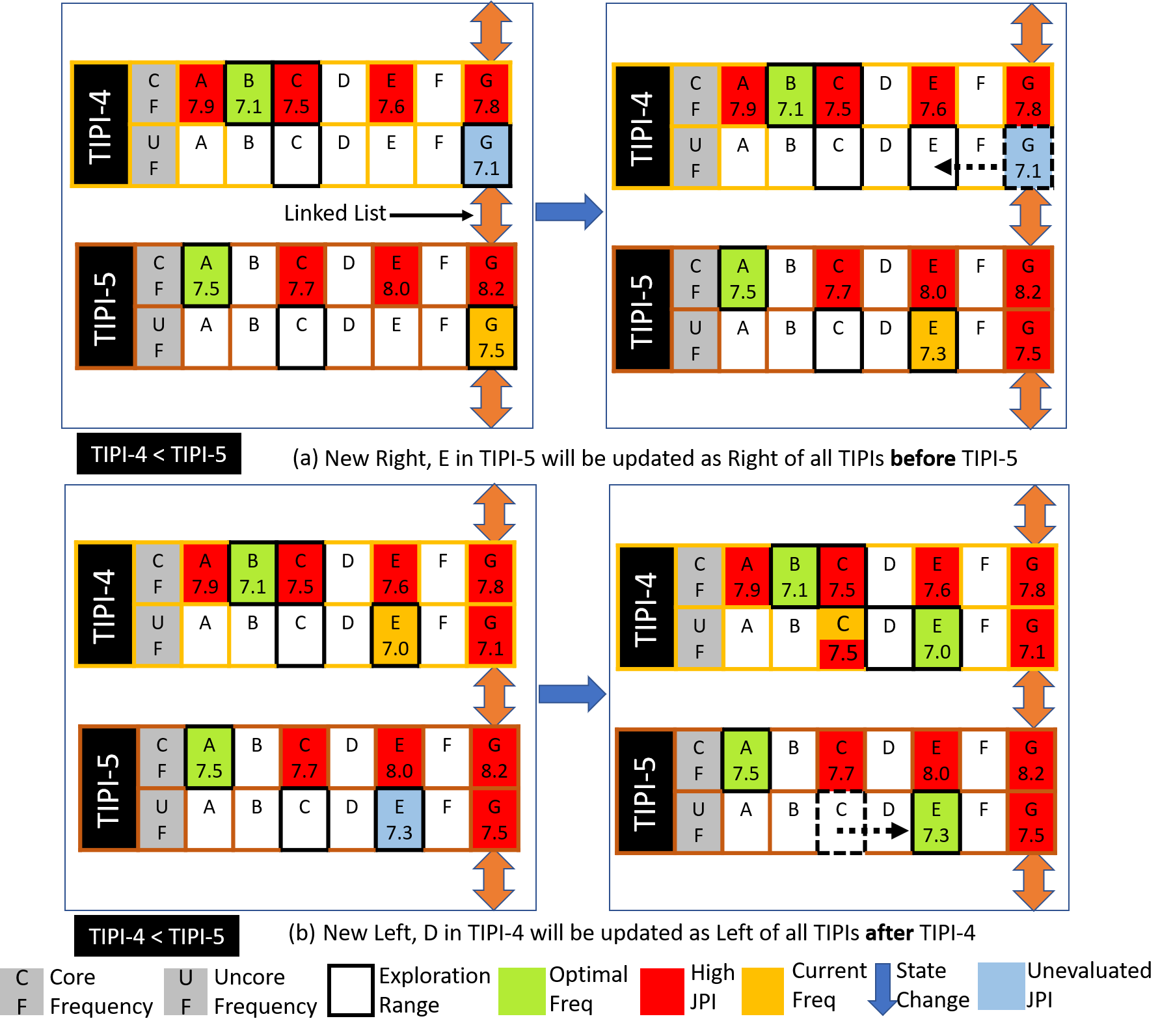}
    \captionof{figure}{Revalidation of \textjava{UF} exploration range}
   \label{fig:impl6}
\end{figure}

Figure~\ref{fig:impl6} demonstrates the same optimization discussed
above, but only for revalidating the uncore exploration range.
In Figure~\ref{fig:impl6}(a), before phase change from left to right,
MAP was set to TIPI-5 with the JPI available for \urb=\ufl{G}. Uncore was
then set to \ufl{E}, and after some \interval, JPI at \ufl{E} was
found to be less than that at \ufl{G} for TIPI-5. This changed
the \urb from \ufl{G} to \ufl{E} at TIPI-5. However, as TIPI-4 is
compute-bound relative to the TIPI-5, its \urb will also shift
from \ufl{G} to \ufl{E}. 

In Figure~\ref{fig:impl6}(b), before phase change from left to right,
MAP was set to TIPI-4 with the JPI available for \urb=\ufl{E}.
Uncore was then set to \ufl{C}, and after some \interval, JPI at \ufl{C}
was found to be more than that at \ufl{E} for TIPI-4. This changes
the \ulb from \ufl{C} to \ufl{D} at TIPI-4. Now, as the \ulb and \urb
at TIPI-4 are consecutive frequencies, \ouf for TIPI-4 will be set to \ufl{E}.
As TIPI-5 is memory-bound relative to the TIPI-4, its \ulb will also shift
from \ufl{C} to \ufl{E} (TIPI-4's \ouf). However, now as both \ulb and \urb
in TIPI-5 points to \ufl{E}, its \ouf will also be set to \ufl{E}.    


\subsection{\cf in distributed computing}

\cf is currently suitable for profiling a single multicore parallel program. 
Hence, one can also use it in MPI+X style distributed computing programs, 
where a single process is executed at each node for inter-node communications (MPI, UPC++, etc.), 
and a multithreaded library (e.g., OpenMP) is used for parallelizing intra-node computations. 
However, \cf cannot regulate the processor frequencies to mitigate the workload imbalance 
between the process~\cite{bhalachandra2017adaptive}. Thereby, its scope is limited to the 
node level parallel regions (e.g., OpenMP) in regular MPI+X 
parallel programs that do not exhibit any load-imbalance due to overlapping computation and communication. 
We aim to extend \cf for achieving energy efficiency in MPI+X style hybrid parallel programs as future work.

\section{Experimental Evaluation}
\label{sec:results}

\begin{figure*}
    \centering
    \subfigure[Energy savings relative to \orig]{\label{graph:ompenergy}\includegraphics[width=\textwidth]{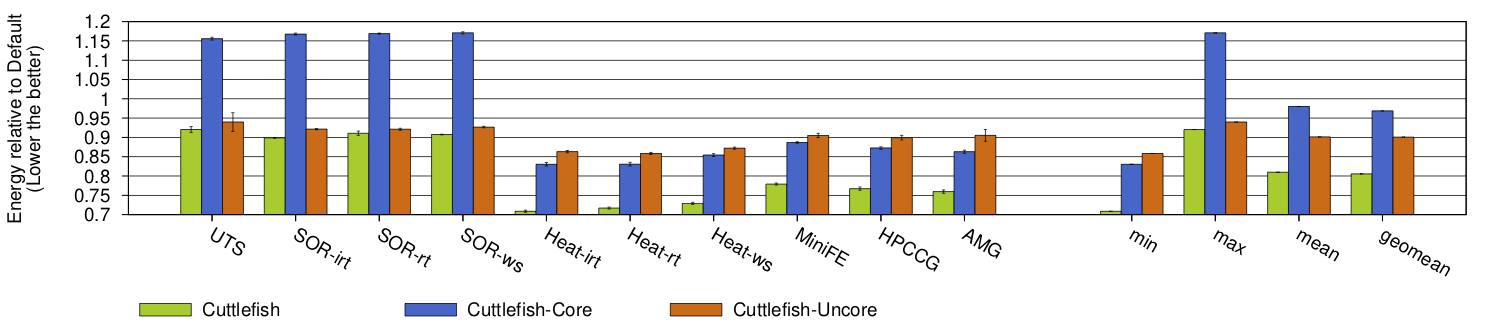} }
    \subfigure[Execution time relative to \orig]{\label{graph:omptime}\includegraphics[width=\textwidth]{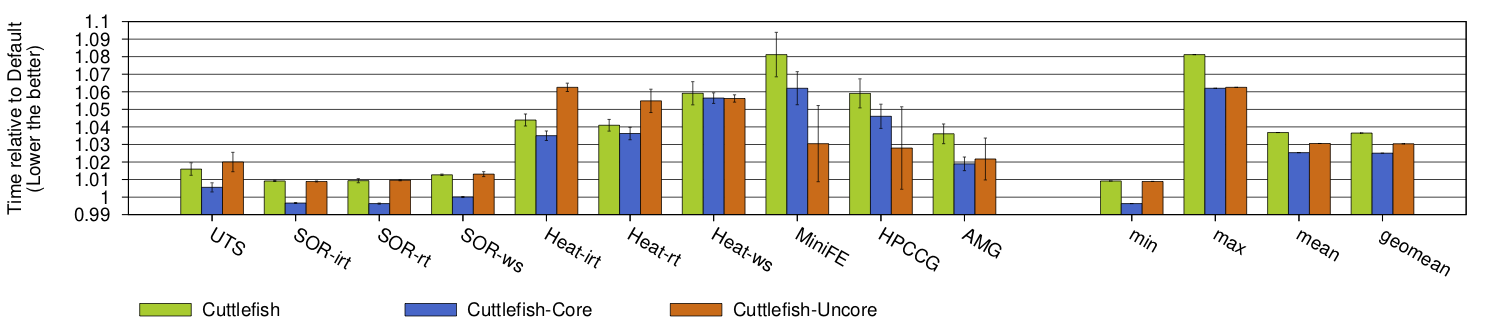} }
    \subfigure[\rebuttal{EDP relative to \orig}]{\label{graph:ompedp}\includegraphics[width=\textwidth]{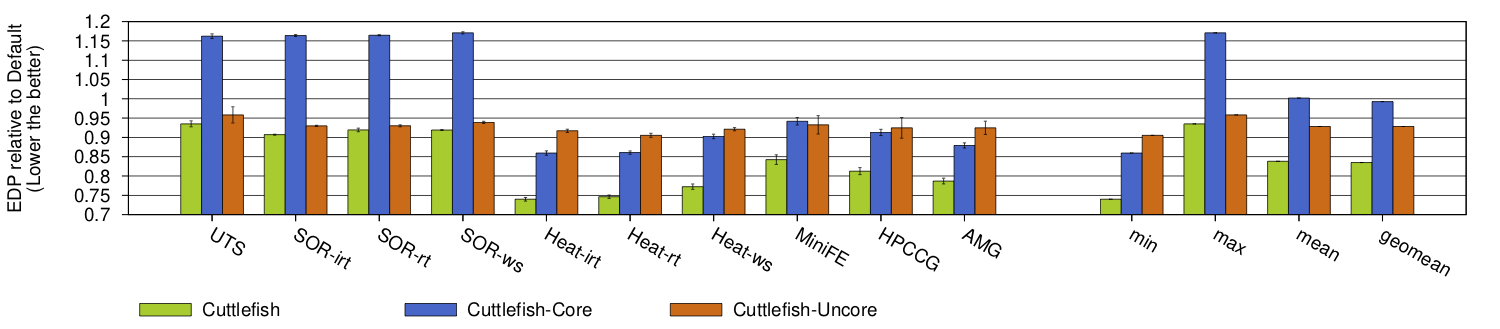} }
    \caption{Experimental evaluation using \omp}
   \label{graph:omp}
\end{figure*}


We provide build-time flags for the three different policies, \cf, \cfc, and \cfu, in the \cf library.  
Section~\ref{sec:impl} described the design and implementation of \cf that dynamically
changes both the core and uncore frequencies. 
\cfc and \cfu are the subsets of the \cf implementation.
\cfc only adapts the core frequencies by fixing the uncore frequency
at its maximum (3.0GHz). \cfu only adjusts the uncore
frequencies by setting the cores frequency at its maximum (2.3GHz).
For the first TIPI-range, both \cfc and \cfu start the frequency exploration 
described in Section~\ref{sec:singletipi} in their respective default 
exploration ranges (1.2GHz-2.3GHz for core and 1.2GHz-3.0GHz
for uncore). For subsequent TIPI ranges,
they use all the runtime optimizations described in Section~\ref{sec:multitipi1}
and Section~\ref{sec:multitipi2} but restrict these optimizations 
to core-only in \cfc and uncore-only in \cfu. 

We begin our evaluation of \cf policies by measuring energy savings and slowdown for both
\omp and \hc benchmarks. We then compare the \ocf and \ouf
calculated by the \cf to \orig for all frequent TIPI ranges.

\subsection{Evaluation of \omp benchmarks}
\label{res:omp}

Figures~\ref{graph:ompenergy},~\ref{graph:omptime}, \rebuttal{and~\ref{graph:ompedp} compare the energy savings,
execution time, and EDP, respectively,} of the \omp benchmarks while using \cf's 
policies to that of the \orig. Recall from
Section~\ref{sec:methodology}, \orig was executed by setting the \textjava{performance} power governor
\rebuttal{that runs each core at the highest frequency (2.3GHz). During the \orig execution,
uncore frequency is controlled by the processor based on the memory access pattern}. 
Geomean energy-savings with \cf, \cfc, and \cfu
are 19.6\%, 3.1\%, and 9.9\%, respectively. 
Geomean loss in performance with \cf, \cfc, and \cfu are 3.6\%, 2.5\%, and 3\%, respectively.
Compared to the \orig, \cfc required more energy in \uts, \sorir, \sorrr, and \sorws.
As these benchmarks are purely compute-bound 
(see TIPI-range in Table~\ref{tab:bench}), \cfc would fix the \ocf 
for these benchmarks at the highest frequency (2.3GHz).
However, unlike \cfc,
the \orig optimizes the uncore frequency, thereby requiring \rebuttal{less}
energy than \cfc. 
\rebuttal{Geomean EDP savings by \cf, \cfc, and \cfu are
16.5\%, 0.7\%, and 7.2\%, respectively, over the \orig.}  

For memory-bound benchmarks 
(Heat variants, \minife, \hpccg, and \amg), energy-savings
from \cfc and \cfu are almost similar 
(the difference is less than 5\%). Although there is some degradation in time, the max is 6.3\% (\heatws). 
This demonstrates that
using \cf even in core-only or uncore-only
mode is also quite effective for memory-bound benchmarks. 
As \cf adapts both core and uncore frequencies,
it has the potential to save energy in both compute and memory-bound benchmarks.
Energy savings are higher in the memory-bound benchmarks because \cf 
can adapt both the uncore and core frequencies.
However, as it has to explore both these cases, the performance degradation
is slightly higher (max 8.1\% in \minife) than
the compute-bound benchmarks (max 1.6\% in \uts).

\subsection{Evaluation of \hc benchmarks}
\label{res:hclib}

\begin{figure*}
    \centering
    \subfigure[Energy savings relative to \orig]{\label{graph:hclibenergy}\includegraphics[width=\textwidth]{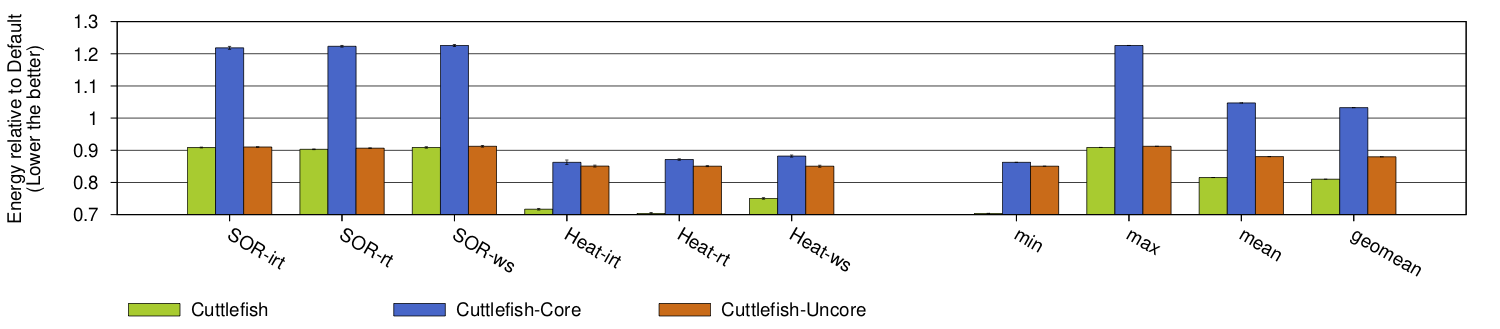} }
    \subfigure[Execution time relative to \orig]{\label{graph:hclibtime}\includegraphics[width=\textwidth]{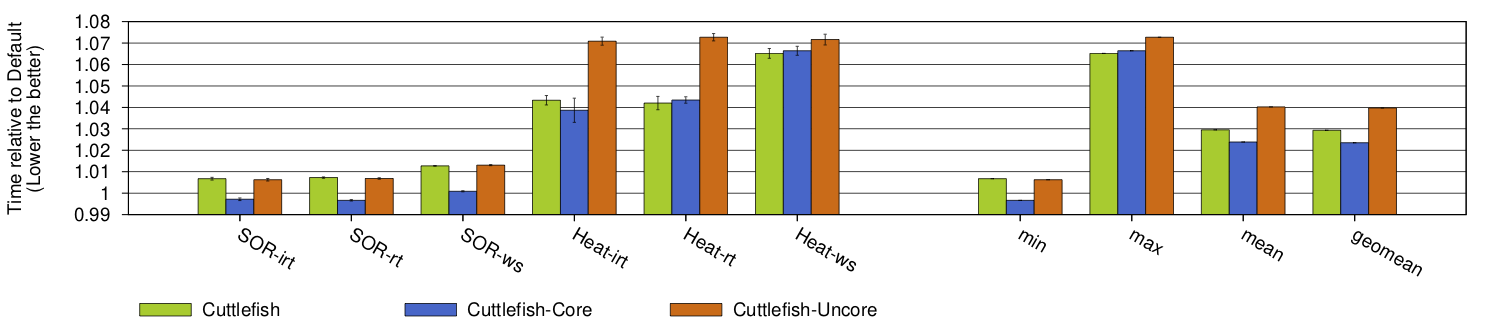} }
    \subfigure[\rebuttal{EDP relative to \orig}]{\label{graph:hclibedp}\includegraphics[width=\textwidth]{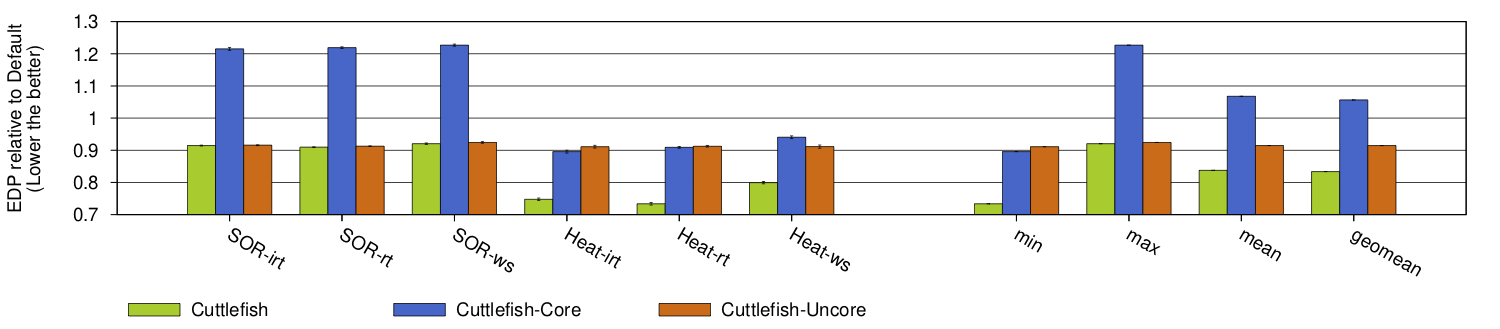} }
    \caption{Experimental evaluation using \hc}
   \label{graph:hclib}
\end{figure*}


%

To support our hypothesis that \cf is a programming model oblivious, we now
present the evaluation of \cf policies using \hc implementations 
of SOR and Heat variants. We omit \minife, \hpccg, and \amg due
to porting challenges. We also discarded \uts as 
this benchmark has an inbuilt work-stealing implementation, and modifying
it to use \hc would alter its algorithm.
\hc supports \afp task parallelism and internally uses a work-stealing runtime
for dynamic load-balancing these asynchronous tasks.
The results of this experiment are shown in Figure~\ref{graph:hclib}. Figure~\ref{graph:hclibenergy}, 
Figure~\ref{graph:hclibtime}, \rebuttal{and Figure~\ref{graph:hclibedp}} 
compare the energy savings, time degradation\rebuttal{, and EDP, respectively,} 
for \cf policies to that of the \orig. Comparing these results with Figure~\ref{graph:omp}
shows that \cf delivers similar results in \hc and \omp benchmarks.

\subsection{Frequency settings using \cf}
\label{res:opt}

\begin{table}[]
\resizebox{\columnwidth}{!}{%
\begin{tabular}{|c|c|c|c|c|c|c|c|}
\hline
\multirow{2}{*}{Benchmark} & \multicolumn{2}{l|}{\begin{tabular}[c]{@{}c@{}}TIPI ranges (\%) \\ having\\ \ocf and \ouf\end{tabular}}            & \multirow{2}{*}{Frequent TIPI Ranges} & \multicolumn{2}{c|}{Cuttlefish} & \multicolumn{2}{c|}{Default} \\ \cline{2-3} \cline{5-8} 
                           & \multicolumn{1}{c|}{\ocf} & \multicolumn{1}{c|}{\ouf} &                             & \ocf              & \ouf            & CF            & \ouf           \\ \hline
UTS                        & 100\%                   & 100\%                   & 0.000- 0.004 (100\%)        & 2.3 ($\pm0\%$)     & 1.3 ($\pm9\%$)    & 2.3           & 2.2          \\ \hline
SOR-irt                    & 100\%                   & 100\%                   & 0.024 -0.028 (100\%)        & 2.3 ($\pm0\%$)      & 1.2 ($\pm0\%$)    & 2.3           & 2.2          \\ \hline
SOR-rt                     & 100\%                   & 100\%                   & 0.024 -0.028 (100\%)        & 2.3 ($\pm0\%$)      & 1.2 ($\pm5\%$)    & 2.3           & 2.2          \\ \hline
SOR-ws                     & 100\%                   & 100\%                   & 0.024 -0.028 (93\%)         & 2.3 ($\pm0\%$)      & 1.2 ($\pm0\%$)    & 2.3           & 2.2          \\ \hline
Heat-irt                   & 50\%                    & 25\%                    & 0.064-0.068 (88\%)          & 1.2 ($\pm0\%$)      & 2.2 ($\pm0\%$)    & 2.3           & 3.0          \\ \hline
\multirow{2}{*}{Heat-rt}   & \multirow{2}{*}{33\%}   & \multirow{2}{*}{33\%}   & 0.060-0.064 (15\%)          & -               & -             & 2.3           & 3.0          \\ \cline{4-8} 
                           &                         &                         & 0.064-0.068 (84\%)          & 1.2 ($\pm0\%$)      & 2.2 ($\pm0\%$)    & 2.3           & 3.0          \\ \hline
Heat-ws                    & 18\%                    & 9\%                    & 0.056-0.060 (88\%)          & 1.3 ($\pm9\%$)      & 2.2 ($\pm1\%$)    & 2.3           & 3.0          \\ \hline
MiniFE                     & 44\%                    & 6\%                     & 0.112-0.116 (76\%)          & 1.3 ($\pm9\%$)      & 2.2 ($\pm1\%$)    & 2.3           & 3.0          \\ \hline
HPCCG                      & 35\%                    & 6\%                     & 0.120-0.124 (76\%)          & 1.3 ($\pm9\%$)      & 2.2 ($\pm1\%$)    & 2.3           & 3.0          \\ \hline
\multirow{2}{*}{AMG}       & \multirow{2}{*}{68\%}   & \multirow{2}{*}{3\%}    & 0.144-0.148 (56\%)          & 1.3 ($\pm11\%$)     & 2.2 ($\pm1\%$)    & 2.3           & 3.0          \\ \cline{4-8} 
                           &                         &                         & 0.148-0.152 (25\%)          & 1.2 ($\pm0\%$)      & 2.2 ($\pm0\%$)    & 2.3           & 3.0          \\ \hline
\end{tabular}
}
\caption{\ocf and \ouf set by \cf in \omp benchmarks and its comparison with the \orig settings}
\label{tab:optimal}
\end{table}


Table~\ref{tab:optimal} shows the percentage of distinct TIPI ranges (see Table~\ref{tab:bench}) 
for which \cf was able to find the \ocf and \ouf in \omp benchmarks. This table also lists the 
\ocf and \ouf set by the \cf for frequently found TIPI ranges and then compare them with the \orig settings.
\amg was having the highest number of distinct TIPI ranges (total 60), due to which it experiences
a wide variation in MAP during the execution timeline (Figure~\ref{graph:timelineTIPI}).
\cf discovered \ocf and \ouf in 68\% and 3\% of the
distinct TIPIs, respectively, even for such an unstable execution. 
We found that the major exploration was carried out only for the frequently
found TIPI ranges (total two). Still, optimal frequencies for the rest were mostly set using the runtime optimizations
described in Section~\ref{sec:multitipi1} and Section~\ref{sec:multitipi2}. As 58 of the 60 distinct 
TIPI ranges appear in less than 10\% of the total \interval durations, \cf
doesn't get too many opportunities to explore \ouf in \amg for these infrequent TIPI ranges (\ocf is explored before \ouf). 

For the frequently found TIPI ranges in each benchmark, \cf 
accurately set both the \ocf and \ouf. These
frequencies match the trend observed during motivational analysis 
in Section~\ref{sec:tipitrend}. \heatrr has only two distinct TIPI ranges. 
Although TIPI 0.060-0.064 appears during 15\% of \interval durations in \heatrr, 
their occurrences are widespread across the execution timeline. Hence, in \heatrr, \cf could not
set \ocf and \ouf for the TIPI range 0.060-0.064.
Due to the performance power governor in \orig, the \textjava{CF} was set
to 2.3GHz (\cfl{max}). 
The processor's uncore settings in \orig were 2.2GHz and 3.0GHz (\ufl{max}) for 
compute-bound and memory-bound benchmarks, respectively. 

\subsection{Sensitivity to \interval}

\begin{table}
\begin{algocolor}
\resizebox{0.6\columnwidth}{!}{%
\begin{tabular}{|c|c|c|} 
\hline
T\_inv & Energy Savings~ & Slowdown   \\ 
\hline
10ms              & 19.5\%          & 4.1\%                    \\ 
\hline
\textbf{20ms}     & \textbf{19.4\%} & \textbf{3.6\%}           \\ 
\hline
40ms              & 18.8\%          & 2.9\%                    \\ 
\hline
60ms              & 17.8\%          & 2.9\%                    \\
\hline
\end{tabular}
}
\caption{\rebuttal{Geomean energy savings and slowdown in \omp benchmarks using different values of \interval in \cf}}
\label{tab:tinv}
\end{algocolor}
\end{table}

\rebuttal{
This section discusses the impact of \interval on the overall energy savings 
and slowdown from using \cf.
Table~\ref{tab:tinv} compares the geomean energy savings and slowdown of \omp benchmarks 
relative to \orig by using \cf at different values of \interval. 
RAPL MSRs are updated every 1 ms on Intel Haswell~\cite{ddcm}. Hence, we chose
\interval as 10$\times$, 20$\times$, 40$\times$, and 60$\times$ of this processor default.
We can observe there is a marginal decrease in energy savings and 
slowdown with increasing \interval.
Recall from Section~\ref{res:omp}, \cf saved
more energy in memory-bound benchmarks (22\%--29\%) than compute-bound (8\%--10.1\%), 
but with some performance degradation (3.6\%--8.1\%). 
This is because optimal frequencies were lower for both core (1.2GHz/1.3GHz) and uncore (2.2GHz) 
in memory-bound benchmarks (Table~\ref{tab:optimal}).
As \cf performs frequency exploration linearly, starting from the 
highest to lowest frequencies, higher \interval increases the exploration time, 
thereby allowing benchmarks to run longer at higher frequencies, 
causing a slight reduction in energy savings and slowdown.
We chose \interval=20ms as the default configuration of \cf,
as it achieved similar energy savings as with \interval=10ms, 
but with lesser slowdown.
}

\subsection{Summary}

\rebuttal{
The encouraging results for \cf demonstrate its efficacy 
for achieving energy efficiency using a combination of DVFS and UFS on modern multicore processors in a wide range of applications with minimal impact on execution time. 
\cf, therefore, promises a one-stop solution as it uses both these dynamically available fine-grain power controls on Intel processors to regulate power.
Its two variants, \cfc and \cfu, provide alternatives to using both DVFS and UFS together. Overall, \cfu achieved better EDP than \cfc. While \cfu is suitable for both compute and memory-bound applications, \cfc is not ideal for compute-bound applications as it sets the uncore to the maximum frequency that is otherwise adaptable in the default settings. However, on memory-bound applications, \cfc was able to deliver EDP close to that of \cfu.
} 

\section{Related Work}
\label{sec:related}

Most energy-efficient HPC research in software has focused on making efficient use of micro-architectural power control support available. 
DVFS is supported by both AMD and Intel processors.
It has been the primary choice to control processor frequency in several studies~\cite{kappiah2005just, hsu2005power, dblp:conf/sc/gefc05, dblp:conf/ppopp/freehl05, dblp:conf/cluster/kimurashbt06, dblp:conf/ics/rountreelssfb09, tiwari12,bhalachandra2017adaptive, bhalachandra2017improving, basireddy2018workload}. 
In the past, processors provided only chip-level DVFS where a frequency change affected all the cores.
Most modern processors today support core-specific implementations of DVFS, allowing each core to operate at different frequencies.
Intel has provided core-specific support for DVFS from its Haswell generation processors.
DDCM has also been used to control processor frequency at a core level in several studies~\cite{sundriyal2014automatic, wang2015evaluating, bhalachandra2015using}.

The target programming model of prior work has dictated much of their design and implementation decisions.
The approaches that target the Message Passing Interface (MPI) applications mainly involve mitigation of workload imbalance between the process (slack)~\cite{dblp:conf/ppopp/freehl05, dblp:conf/cluster/kimurashbt06, dblp:conf/ics/rountreelssfb09, bhalachandra2017adaptive}.
Other MPI-centric solutions address cases where the processor cores wait on the memory or network~\cite{kandalla2010designing, vishnu2010designing, sundriyal2011per, hoefler2014energy, venkatesh2015case, bhalachandra2017improving}.
Concurrency throttling has been widely used by adapting the thread count in OpenMP programs that are memory-constrained to reduce power consumption~\cite{curtis2006online, curtis2008prediction, li2010hybrid, porterfield2013openmp}.
Methods~\cite{wang2015using} that involve assigning optimal frequencies to OpenMP loops based on their memory access patterns have been proposed.
Static analysis of loop-based parallel programs has been used to compute each loop chunk's workload 
and then tune the chip frequency dynamically by calculating the remaining workload~\cite{SRNV2017}. 
Iterative parallel programs provide an opportunity to dynamically tune the frequencies of the cores according to
the workload information of the tasks collected with the online profiling of the first few iterations~\cite{6969445}.
The impact of power capping on compiler transformations has been studied by utilizing the memory 
access density information of OpenMP loops~\cite{wang2016performance}. 
Fine-tuning the core frequencies based on online profiling of thief-victim relationships and the size of deques in 
a work-stealing runtime is another approach for programming model-specific solution~\cite{RHLY2014}. 
\rebuttal{
There have been attempts to standardize the interfaces controlling power at 
different levels of the HPC system hierarchy. GEOPM~\cite{geopm} is one such
implementation that leverages feedback from the application to identify which 
nodes are on the critical path and then adjusts processor power cap settings 
to accelerate the critical path and improve the application's time-to-solution.
The current effort may easily fit as a part of the third-party software components that GEOPM and such standardized interfaces support.
}

UFS is a recent research topic for achieving energy efficiency, although it 
has been available to the user since the Intel Haswell processor generation~\cite{ufs}.
The high efficacy of UFS has been demonstrated in terms of its energy-saving potential~\cite{sundriyal2018comparisons}.
UFS has been used to dynamically adapt the uncore frequency based on the DRAM
power usage~\cite{GNMF2019}. A machine learning-based UFS model 
has been proposed to gather performance metrics from offline executions
and predict the optimal uncore frequency~\cite{BSAB2019}.
The work described in~\cite{sundriyal2016joint} aims to achieve energy efficiency by using both DVFS and \rebuttal{DRAM scaling}. However, unlike \cf, this implementation needs prior data to learn some machine-dependent parameters using regression analysis. These parameters are used at runtime to calculate the optimal core and uncore frequency.

\section{Conclusion}
\label{sec:conclusion}

In this paper, a programming model oblivious C/C++ library
is proposed for achieving energy efficiency that is
not limited by language constraints and semantics.
The model can detect the changes in the memory access patterns even for the same application across multiple 
language implementations to assign optimal frequencies.
Using both DVFS and UFS allows additional energy savings than solutions that utilize a single power control.
The effectiveness of the current approach is demonstrated for irregular-tasking as well as work-sharing pragmas.
The use of both micro-kernels and real-world HPC mini-applications 
for evaluation helps better understand the library's efficacy for production use.

In the future, we would like to extend this framework to support hybrid MPI+X applications.
With the emergence of scientific workflows in HPC systems, we want to explore the possibility of using \cf
to control the power of co-running components of a workflow on a node.
Graphics processing units (GPUs) are ever more prevalent in HPC today. 
The energy-efficiency potential in the context of GPU-offload support offered by many high-level languages is another area worth exploring.
\balance

\section*{Acknowledgments}
The authors are grateful to the anonymous reviewers for their suggestions on improving the presentation of the paper.
Results presented in this paper were obtained using the Chameleon testbed 
supported by the National Science Foundation. This research used resources of the National 
Energy Research Scientific Computing Center, a DOE Office of Science 
User Facility supported by  the Office of Science of the U.S. Department of 
Energy under Contract No. DE-AC02-05CH11231.



\end{document}